\documentclass[prd,preprint,superscriptaddress,preprintnumbers,eqsecnum,showpacs,nofootinbib,nobibnotes]{revtex4-1}
\usepackage{amsfonts,amsmath,amssymb,bm}
\usepackage[table]{xcolor}
\usepackage{graphicx} 
\usepackage{mathtools}
\usepackage{boondox-cal}
\usepackage{xcolor}

\usepackage[utf8]{inputenc}
\usepackage{slashed}

\newcommand{\be}{\begin{equation}}
\newcommand{\bea}{\begin{eqnarray}}
\newcommand{\ba}{\begin{align}}
\newcommand{\ee}{\end{equation}}
\newcommand{\eea}{\end{eqnarray}}
\newcommand{\ea}{\end{align}}
\newcommand{\cw}{\cellcolor{white}}

\definecolor{zero2}{rgb}{0.88,0.88,.88}

\def\1eq#1{Eq.~(\ref{#1})}

\def\2eqs#1#2{Eqs.~(\ref{#1}) and~(\ref{#2})}
\def\3eqs#1#2#3{Eqs.~(\ref{#1}),~(\ref{#2}) and~(\ref{#3})}
\def\4eqs#1#2#3#4{Eqs.~(\ref{#1}),~(\ref{#2}),~(\ref{#3}) and~(\ref{#4})}

\def\tbarc{\bar{\mathcal{c}}^*}
\def\tT{\mathcal{T}_1}
\def\tU{\mathcal{U}}
\def\tR{\mathcal{R}}

\def\rc#1{\rho_{#1}}
\def\trc#1{{\tilde \rho}_{#1}}

\def\op#1{{\cal O}_{#1}}
\def\lop#1{{\cal I}_{#1}}

\def\s#1{{\scriptscriptstyle #1}}

\def\G{\Gamma}

\def\T{T_1}

\def\s{\mathcal{s}}

\def\hphi0{{\hat\phi}_0}

\def\d{\!\mathrm{d}^4x\,}

\def\cfxt#1{\theta_{#1}}

\def\cfgi#1{\lambda_{#1}}
\def\cfgit#1{{\tilde \lambda}_{#1}}
\def\cfps#1{\vartheta_{#1}}

\makeatletter
\def\user@resume{resume}
\def\user@intermezzo{intermezzo}
\newcounter{previousequation}
\newcounter{lastsubequation}
\newcounter{savedparentequation}
\renewenvironment{subequations}[1][]{%
      \def\user@decides{#1}%
      \setcounter{previousequation}{\value{equation}}%
      \ifx\user@decides\user@resume 
           \setcounter{equation}{\value{savedparentequation}}%
      \else  
      \ifx\user@decides\user@intermezzo
           \refstepcounter{equation}%
      \else
           \setcounter{lastsubequation}{0}%
           \refstepcounter{equation}%
      \fi\fi
      \protected@edef\theHparentequation{%
          \@ifundefined {theHequation}\theequation \theHequation}%
      \protected@edef\theparentequation{\theequation}%
      \setcounter{parentequation}{\value{equation}}%
      \ifx\user@decides\user@resume 
           \setcounter{equation}{\value{lastsubequation}}%
         \else
           \setcounter{equation}{0}%
      \fi
      \def\theequation  {\theparentequation  \alph{equation}}%
      \def\theHequation {\theHparentequation \alph{equation}}%
      \ignorespaces
}{%
  \ifx\user@decides\user@resume
       \setcounter{lastsubequation}{\value{equation}}%
       \setcounter{equation}{\value{previousequation}}%
  \else
  \ifx\user@decides\user@intermezzo
       \setcounter{equation}{\value{parentequation}}%
  \else
       \setcounter{lastsubequation}{\value{equation}}%
       \setcounter{savedparentequation}{\value{parentequation}}%
       \setcounter{equation}{\value{parentequation}}%
  \fi\fi
  \ignorespacesafterend
}
\def\CT@@do@color{%
	\global\let\CT@do@color\relax
		\@tempdima\wd\z@
		\advance\@tempdima\@tempdimb
		\advance\@tempdima\@tempdimc
		\advance\@tempdimb\tabcolsep
		\advance\@tempdimc\tabcolsep
		\advance\@tempdima1.5\tabcolsep
	\kern-1.5\@tempdimb
	\leaders\vrule
	\hskip\@tempdima\@plus  1fill
	\kern-1.5\@tempdimc
	\hskip-\wd\z@ \@plus -1fill }
\makeatother

\begin{document}

\title{
Off-shell renormalization\\ in the presence of dimension~6 derivative operators.\\ 
III. Operator mixing and $\beta$ functions}

\date{January 21, 2020}

\author{D. Binosi}
\email{binosi@ectstar.eu}
\affiliation{European Centre for Theoretical Studies in Nuclear Physics
and Related Areas (ECT*) and Fondazione Bruno Kessler, Villa Tambosi, Strada delle Tabarelle 286, I-38123 Villazzano (TN), Italy}

\author{A. Quadri}
\email{andrea.quadri@mi.infn.it}
\affiliation{INFN, Sezione di Milano, via Celoria 16, I-20133 Milano, Italy}

\begin{abstract}
\noindent
We evaluate the one-loop $\beta$ functions of all dimension 6 parity-preserving operators in the Abelian Higgs-Kibble model. No on-shell restrictions are imposed; and the (generalized) non-polynomial field redefinitions arising at one-loop order are fully taken into account. The operator mixing matrix is also computed, and its cancellation patterns explained as a consequence of the functional identities of the theory and power-counting conditions.
\end{abstract}


\maketitle

\section{Introduction}

New physics beyond the Standard Model (SM) can be characterized in a model independent and systematic fashion within the Effective Field Theories (EFTs) framework, in which the (renormalizable) tree-level SM action is supplemented with the terms ($k\geq5$)
\begin{align}
    S_0^{[k]}=\int\!{\mathrm{d}^4x}\,\sum_i c^{[k]}_i{\cal O}^{[k]}_i,
\end{align}
where ${\cal O}^{[k]}_i$ are $k$-dimensional operators whose dimension  dictates the suppression of the corresponding coefficients $c_i^{[k]}$ in terms of powers of a high-energy scale $\Lambda$. The resulting Standard Model Effective Field Theory (SMEFT) action 
\begin{align}
	S_0 \equiv \underbrace{\int\!{\mathrm{d}^4x}\,{\cal L}_{\mathrm{SM}}}_{\sum_{k=2}^4 S_0^{[k]}}+\sum_{k\geq5} S_0^{[k]},
\end{align}
is not renormalizable in the usual (power counting) sense; it is, nevertheless, renormalizable in the modern sense~\cite{Gomis:1995jp}, as all the divergences can be cancelled through the renormalization of the (infinite) number of terms in the bare action while respecting the symmetries of the theory.

When addressing operator mixing in such theories on-shell calculations are sufficient. Indeed while it has been known since a long time that there is ultraviolet (UV) mixing between gauge invariant and gauge variant (unphysical) operators (also known as `alien' operators~\cite{Collins:1994ee}), it has also been shown that such mixing can be made to vanish by a suitable choice of the basis in the space of local operators~\cite{Joglekar:1975nu, KlubergStern:1974rs,KlubergStern:1974xv,KlubergStern:1975hc}; additionally, alien operators have been shown to be cohomologically trivial and therefore have vanishing on-shell correlators~\cite{Joglekar:1975nu} (for a review see also \cite{Barnich:2000zw}). This fact is at the basis of recent computations in the literature~\cite{Jenkins:2013zja,Jenkins:2013zja,Jenkins:2013wua,deFlorian:2016spz,Brivio:2017vri} as it implies that for certain purposes, {\it e.g.}, when evaluating anomalous dimensions and/or $S$-matrix elements, one can consider only on-shell inequivalent operators~\cite{Grzadkowski:2010es}.

A separate issue, however, is the evaluation of the $\beta$-functions of the theory. For this purpose one needs to extend the approach adopted in the power-counting renormalizable case~\cite{Machacek:1983tz,Machacek:1983fi,Machacek:1984zw} to EFTs; in particular, one must work out a procedure to fix the generalized field redefinitions (GFRs) that do arise in these models. Here `generalized' means that, at variance with the power-counting renormalizable case, these redefinitions are not linear in the quantum fields (in fact,
not even polynomial already at one-loop order, as we will show). The matching of the couplings order by order in the loop expansion, once the GFRs' effects are taken into account, is the next technical step required to match the model with its UV completions while respecting the locality of the low energy theory also at higher loop orders, since it allows to unequivocally fix the correct counter-terms needed to subtract
overlapping divergences with local counter-terms.

To attain these goals, in~\cite{Binosi:2019olm} it has been developed a general theory for the recursive subtraction of off-shell UV divergences order by order in the loop expansion applicable to EFTs displaying a spontaneously broken symmetry phase. This is achieved by solving the Slavnov-Taylor (ST) identity to all-orders, which allows in turn to disentangle the gauge-invariant contributions to the off-shell one-particle irreducible (1PI) amplitudes from those associated with the gauge fixing and field redefinitions, which, in a general EFT, can be (and indeed are) non polynomial (and cannot obviously be accessed staying on-shell). Next, in~\cite{Binosi:2019nwz} this algebraic technique has been applied to study the Abelian Higgs-Kibble (HK) model in the presence of the dimension 6 operator $(g/\Lambda)\phi^\dagger\phi(D_\mu\phi)^\dagger D^\mu\phi$, which, giving rise to an infinite number of one-loop divergent diagrams, maximally violates power counting. In particular, the complete renormalization of all the radiatively generated dimension 6 operators has been carried out together with the determination of the full $g$-dependence of the $\beta$-function coefficients. 

Before moving on to consider the full dimension 6 SMEFT~\cite{in-prep}, there is just one aspect that has been left out in the study of its Abelian sibling: namely, the analysis of the full off-shell renormalization when all inequivalent parity-preserving dimension 6 operators (classified according to~\cite{Grzadkowski:2010es}) are added to the power counting renormalizable action. And this constitutes precisely the subject of the present paper. 

From the point of view of the EFT renormalization programme of~\cite{Gomis:1995jp}, what we achieve here is to fully evaluate all the terms appearing in the renormalized action $S$ at one loop (in the relevant sector of dimension $\leq 6$), expressed as
\begin{align}
    S_0 = S + \hbar \Delta_{1} + \cdots. 
    \label{bare.ren.act}
\end{align}
At zero antifields, $\Delta_1$ collects  one-loop gauge-invariant counterterms. The renormalized action has the same form as the original bare action $S_0$; in particular, it can be expanded on a basis of gauge-invariant operators (in the zero antifield sector). However, these counterterms are not enough to renormalize the theory: one must also take into account the effects of GFRs, that are implemented according to a canonical transformation with respect to the Batalin-Vilkovisky (BV) bracket associated with the gauge symmetry of the model~\cite{Gomis:1995jp}. The transformed bare action $S'_0$ takes then the form
\begin{align}
	S'_0= S +\hbar\left[\Delta_1+ (F_1, S)\right] + \cdots,
	\label{weinberg.ren}
\end{align}
where $F_1$  is the one loop term in the loop expansion $F(t)=\hbar t F_1+\cdots$ of the generator of the canonical transformation responsible for the field-antifield redefinition: 
$\Phi \rightarrow \Phi'(\Phi,\Phi^*)$, $\Phi^* \rightarrow \Phi^{*'}(\Phi,\Phi^*)$ on $S'_0[\Phi',\Phi^{'*}] = S_0[\Phi,\Phi^{*}]$.
Being canonical, this transformation preserves the fundamental BV brackets
$(\Phi^{'i},\Phi^{'*}_j) = \delta^i_j \,, \quad 
(\Phi^{'i},\Phi^{'j}) = (\Phi^{'*}_i, \Phi^{'*}_j) = 0,$
and is obtained by solving the differential equation $\dot{S}_0(t) = ( F(t), S_0(t))$ with the boundary condition \mbox{$S_0(0)=S_0$}, see~\cite{Gomis:1995jp}. Such canonical transformation generalizes the usual linear wave function renormalizations of the power-counting renormalizable cases. It plays a crucial and ubiquitous role in the SMEFT renormalization program, as we will show.

The paper is organized as follows. In Sect.~\ref{sec.not} we set up our notation and, in order to make the work self-contained  we briefly review the most salient features of the $X$-formalism. Then, in  Sects.~\ref{sec.oneloop.divs} and~\ref{sec.dim6.coeffs} the parameterization of the one-loop UV divergences both in the $X$- and the target (original) theory is presented and the mapping between the two theory's formulations derived. GFRs are studied and their form explicitly obtained in Sect.~\ref{sec.gfrs}, whereas the renormalization of dimension 6 gauge invariant operators in the $X$-theory is explicitly carried out in Sect.~\ref{sec.ren}. Finally, in Sect.~\ref{sec.oneloop.mix} we describe the one-loop mixing matrix in the original theory and compare our results with the literature. Conclusions are presented in Sect.~\ref{sec.concls}. A number of technical issues are discussed in a set of Appendices presented at the end of the paper: functional identities of the $X$-theory and the propagators in Appendices~\ref{app:funct.ids}, \ref{app.beq} and \ref{app:propagators}; the list of gauge invariant operators in Appendix~\ref{app:g.inv.ops}; and, finally, the on-shell operator reduction relations in Appendix~\ref{app:reduction}.

\vfill

\section{Notations and conventions}\label{sec.not}

In the $X$-formalism approach of~\cite{BQ:2019a}, the tree-level vertex functional takes the form
\begin{align}
	\G^{(0)} & = 
	   \int \!\mathrm{d}^4x \, \Big [ -\frac{1}{4} F^{\mu\nu} F_{\mu\nu} + (D^\mu \phi)^\dagger (D_\mu \phi) - \frac{M^2-m^2}{2} X_2^2 - \frac{m^2}{2v^2} \Big ( \phi^\dagger \phi - \frac{v^2}{2} \Big )^2 \nonumber \\
	& - \bar c (\square + m^2) c + \frac{1}{v} (X_1 + X_2) (\square + m^2) \Big ( \phi^\dagger \phi - \frac{v^2}{2} - v X_2 \Big ) \nonumber \\
	&  +\frac{z}{2} \partial^\mu X_2 \partial_\mu X_2 + \frac{g_1 v}{\Lambda^2} X_2 (D^\mu \phi)^\dagger (D_\mu \phi) 
	+  \frac{g_2 v}{\Lambda^2} X_2 F_{\mu\nu}^2 
	+  \frac{g_3 v^3 }{6 \Lambda^2} X_2^3 
	\nonumber \\
	& + \T(D^\mu \phi)^\dagger (D_\mu \phi) +
	U F_{\mu\nu}^2 + R X_2^2
	\nonumber \\
	& + \frac{\xi b^2}{2} -  b \Big ( \partial A + \xi e v \chi \Big ) + \bar{\omega}\Big ( \square \omega + \xi e^2 v (\sigma + v) \omega\Big ) \nonumber \\
	&  + \bar c^* \Big ( \phi^\dagger \phi - \frac{v^2}{2} - v X_2 \Big ) + \sigma^* (-e \omega \chi) + \chi^* e \omega (\sigma + v) \Big ].
	\label{tree.level}
\end{align}
In the expression above, the first line represents the action of the Abelian HK model in the $X$-formalism, where the usual scalar field $\phi \equiv \frac{1}{\sqrt{2}} (\phi_0 + i \chi) = \frac{1}{\sqrt{2}} (\sigma + v + i \chi)$ with $v$ the vacuum expectation value (vev) is supplemented with a singlet field $X_2$, that provides a gauge-invariant parametrization of the physical scalar mode. Notice also that we defined $\phi_0 = \sigma + v$ with $\sigma$ having a zero vev. The field $X_1$ plays instead the role of a Lagrange multiplier: when going on-shell with this field one recovers the constraint\footnote{Going on-shell with $X_1$ yields the condition
$$
    (\square + m^2) \Big (
    \phi^\dagger \phi - \frac{v^2}{2} - v X_2 \Big ) = 0,
$$
so that the most general solution is
$X_2 = \frac{1}{v} \Big (
    \phi^\dagger \phi - \frac{v^2}{2} \Big ) + \eta,$ $\eta$
being a scalar field of mass $m$. However, in perturbation theory the correlators of the mode $\eta$ with any gauge-invariant operators vanish~\cite{Binosi:2019olm}, so that one can safely set $\eta =0$.}
$X_2 \sim \frac{1}{v} ( \phi^\dagger \phi - v^2/2)$, which once inserted back into the first line of Eq.~(\ref{tree.level}), cancels the $m^2$-term leaving the usual Higgs quartic potential with coefficient $\sim M^2/2v^2$. Hence, Green's functions in the target theory\footnote{We define as `target' theory the original theory defined in terms of conventional fields.} have to be $m^2$-independent, a fact that provides a very strong check of the computations, due to the ubiquitous presence of~$m^2$ both in Feynman amplitudes as well as invariants.

The $X_{1,2}$-system comes together with a {\em constraint} BRST symmetry, ensuring that the number of physical degrees of freedom in the scalar sector remains unchanged in the $X$-formalism with respect to the standard formulation relying only on the field $\phi$~\cite{Quadri:2006hr,Quadri:2016wwl}. More precisely, the vertex functional (\ref{tree.level}) is invariant under the following BRST symmetry:
\begin{align}
	\s X_1 = v c; \, \quad \s \phi = \s X_2 = \s c  = 0; \quad \s \bar c = \phi^\dagger\phi - \frac{v^2}{2} - v X_2. 
	\label{u1.brst}
\end{align}
The associated ghost and antighost fields $c, \bar c$ are free. The constraint BRST differential $\s$ anticommutes with the (usual) {\it gauge group} BRST symmetry of the classical action after the gauge-fixing introduced in the fifth line of Eq.~(\ref{tree.level}):
\begin{align}
	s A_\mu &= \partial_\mu \omega;& s \omega &= 0; &  s \bar{\omega} &= b;&  s b &=0 ;&  s \phi &= i e \omega \phi.  
\end{align}
Here $\omega$ ($\bar \omega$) is the U(1) ghost (antighost); the latter field is paired into a BRST doublet with the Lagrange multiplier field $b$, enforcing the usual $R_\xi$ gauge-fixing condition
\begin{align}
    {\cal F}_\xi = \partial A + \xi ev \chi,
    \label{g.f.cond}
\end{align}
with $\xi$ the gauge fixing parameter.

The two BRST symmetries can both be lifted to the corresponding ST identities at the quantum level, provided one introduces a suitable set of so-called  antifields, {\it i.e.}, external sources coupled to the relevant BRST transformations that are non-linear in the quantized fields. The antifield couplings are displayed in the last line of Eq.~(\ref{tree.level}); the ST identities are instead summarized in Appendix~\ref{app:funct.ids}.

The third line of Eq.~(\ref{tree.level}) contains the  dimension 6 parity preserving subset of the gauge-invariant operators described in~\cite{Grzadkowski:2010es}, modulo for the fact that  we use the zero expectation value combination $\phi^\dagger \phi - \frac{v^2}{2} \sim v X_2$ instead of $\phi^\dagger \phi$. We thus see that the classical power-counting renormalizable action is  supplemented in the $X$-formalism by the $X_2$-dependent operators\footnote{In the spirit of~\cite{Grzadkowski:2010es} we drop operators that are on-shell equivalent, {\it i.e.}, that differ by terms vanishing once the classical equations of motion are imposed.}
\begin{subequations}
\begin{align}
    \op{1}^{[6]} &= \int \d  \, F_{\mu\nu}^2 \Big ( \phi^\dagger \phi - \frac{v^2}{2} \Big ) \sim \int \d  \,  v X_2 F_{\mu\nu}^2 ,\\
    \op{2}^{[6]} &= \int \d  \,  \Big ( \phi^\dagger \phi - \frac{v^2}{2} \Big )^3 \sim \int \d  \, v^3 X_2^3, \\
    \op{3}^{[6]} &= \int \d  \,  \Big ( \phi^\dagger \phi - \frac{v^2}{2} \Big ) \square \Big ( \phi^\dagger \phi - \frac{v^2}{2} \Big )
    \sim \int \d  \, v^2 X_2 \square X_2 , \\
    \op{4}^{[6]} &= \int \d  \,  \Big ( \phi^\dagger \phi - \frac{v^2}{2} \Big ) (D^\mu \phi)^\dagger D_\mu \phi \sim  \int \d  \, v X_2 (D^\mu \phi)^\dagger D_\mu \phi.
\end{align}
\end{subequations}
Notice that the operator $\op{3}^{[6]}$ is special in the sense that it does not give rise in the $X$-theory to new interaction vertices: rather it modifies the propagator of the $X_2$-field by rescaling the $p^2$-term~\cite{Quadri:2016wwl} (the full set of propagators of the model is summarized in Appendix~\ref{app:propagators}). Notice also that in comparison with the conventions of~\cite{Binosi:2019olm,Binosi:2019nwz} we have rescaled the higher dimensional coupling constants by a factor $v/\Lambda$ in order to obtain, when mapping back to the target theory, the standard $1/\Lambda^2$ pre-factor for dimension 6 operators.

To maintain a detailed comparison with~\cite{Gomis:1995jp}, we provide in the following some technical details. 

The relevant BV bracket is the one associated with the gauge symmetry, the constraint BRST symmetry invariance being exhausted in the $X_1$-equation, as shown in Appendix~A, see Eqs.~(\ref{sti.c}) and~(\ref{X1.eq}). Next, as the gauge group is Abelian: there is no ghost antifield, since  $s \omega = 0$; the BRST transformation of the gauge field is linear in the quantized fields and thus there is no need to introduce the gauge antifield $A^*_\mu$ for controlling  quantum corrections\footnote{This latter fact can be easily understood since the coupling $$\int \d \, A^*_\mu s A^\mu  = \int \d \, A^*_\mu \partial^\mu \omega$$ does not generate any interaction vertex involving $A^*_\mu$, due to the aforementioned linearity of the BRST transformation of  $A_\mu$ in the quantum fields.} (although algebraically one is allowed to). Also, in the  $R_\xi$-gauge that we employ, there is no need to introduce the antifield $\bar \omega^*$, coupled to the Nakanishi-Lautrup field $b = s \bar \omega$: in fact, see Appendix~B, the $b$-equation~(\ref{b.eq}) and the antighost equation~(\ref{antigh.eq}) imply that at the quantum level there is no dependence on the field $b$ and moreover that the antighost dependence can be reabsorbed by the antifield redefinition~(\ref{chistar.redef}). On the other hand, in the formulation of~\cite{Gomis:1995jp}, where one introduces both $\bar \omega^*$ and $A^*_\mu$, the antighost-dependent sector of the action is recovered from the antifield couplings $\int \d \,( A^*_\mu sA^\mu + \chi^* s\chi )$ via a canonical transformation  with fermionic generator $F = \int \d {\cal F}_\xi \bar \omega$ (that incidentally exactly yields the antifield redefinition in Eq.~(\ref{chistar.redef})). Thus, the dimension $\leq 6$ sector of $S_0$ is
\begin{align}
\sum_{k=1}^6
\left .S_0^{[k]} \right |_{A^*_\mu=\bar \omega^* =0} \equiv
\left . \G^{(0)} \right |_{b=\bar \omega = 0}.
\label{weinberg.map}
\end{align}

At one loop order further operators will be radiatively generated starting from $\G^{(0)}$. Those operators can be however expressed in the target theory as gauge invariant polynomials in the field $\phi$, its (symmetrized) covariant derivatives, the field strength and its ordinary derivatives. This set of variables is particularly suited in order to obtain the coefficients of the one loop invariants controlling the UV divergences of the theory~\cite{Barnich:2000zw}. Additionally, some of these operators will be on-shell equivalent; the reduction to on-shell independent operators is carried out in some detail in Appendix~\ref{app:reduction}.

Returning to~\1eq{tree.level}, we notice that the terms in the third line of Eq.(\ref{tree.level}) respect both BRST symmetries and thus they do not violate either the $X_1$-equation~(\ref{X1.eq}) or the ST identity~(\ref{sti}). Finally, in the fourth row we have added the external sources $T_1, R, U$ required to define the $X_2$-equation at the quantum level in the presence of  additional non power-counting renormalizable interactions, see Eq.(\ref{X2.eq}).

\section{One-loop UV Divergences}\label{sec.oneloop.divs}

In this section we will work out the parameterization of the one-loop UV divergences in the $X$-theory for all the operators giving rise to contributions to  dimension $6$ operators in the target theory.

In what follows subscripts denote functional differentiation with respect to fields and external sources. Thus, amplitudes will be denoted as, {\it e.g.}, $\G^{(1)}_{\chi\chi}$, meaning
\begin{align}
    \G^{(1)}_{\chi\chi} \equiv \left . 
    \frac{\delta^2 \G^{(1)}}{\delta \chi(-p) \delta \chi(p)} \right |_{p=0}.
\end{align}
A bar denotes the UV divergent part of the corresponding amplitude in the Laurent expansion around $\epsilon=4-D$, with $D$ the space-time dimension. Dimensional regularization is always implied, with amplitudes evaluated by means of the packages {\tt FeynArts} and {\tt FormCalc}~\cite{Hahn:2000kx,Hahn:2000jm}. As already remarked, all amplitudes will be evaluated in the Feynman ($\xi=1$) and Landau ($\xi=0$) gauge; this will allow to explicitly check the gauge cancellations in gauge invariant operators and in particular, as we will see, the crucial role of the GFRs in ensuring the gauge independence of ostensibly gauge invariant quantities.

Consider now the UV divergent contributions to one-loop amplitudes. They form a local functional (in the sense of formal power series)  denoted by $\overline{\G}^{(1)}$. Since~$\overline{\G}^{(1)}$ belongs to the kernel of the linearized ST operator ${\cal S}_0$ defined in~\1eq{S0}, {\it i.e.},
\begin{align}
    {\cal S}_0(\overline{\G}^{(1)}) =0,
    \label{uv.div.1loop.st}
\end{align}
the nilpotency of ${\cal S}_0$ ensures that $\overline{\G}^{(1)}$ is the sum of a gauge-invariant functional ${\overline{{\cal I}}}^{(1)}$ and a cohomologically trivial contribution ${\cal S}_0(\overline{Y}^{(1)})$:
\begin{align}
    \overline{\G}^{(1)} =
    {\overline{{\cal I}}}^{(1)}_\mathrm{gi} + {\cal S}_0(\overline{Y}^{(1)}),
    \label{uv.div.1loop.vf}
\end{align}
with GFRs described by the cohomologically trivial term ${\cal S}_0(\overline{Y}^{(1)})$. Eq.~(\ref{uv.div.1loop.vf}) bears in fact a close resemblance with Eq.~(\ref{weinberg.ren}), as, for the model at hand, we find the identifications
\begin{align}
    \Delta_1&=-\left. {\overline{{\cal I}}}^{(1)}_\mathrm{gi}\right|_{b=\bar\omega=0};&
    (F_1,S)&=-{\cal S}_0(\overline{Y}^{(1)}).
\end{align}

Ultimately, we are interested in the UV divergences of dimension 6 gauge invariant operators in the target theory. To identify the invariants in the $X$-theory contributing to these operators the mapping function from the $X$- to the target theory is needed. As explained in~\cite{Binosi:2019olm,Binosi:2019nwz} this amounts to solving the $X_{1,2}$-equations in the $X$-theory via the replacements in Eq.~(\ref{nth.X12.eqs}) and then going on-shell with $X_{1,2}$. At the one loop level it is sufficient to impose the classical equations of motions for $X_{1,2}$. The $X_1$-equation gives
\begin{align}
    X_2 = \frac{1}{v} \Big ( \phi^\dagger \phi - \frac{v^2}{2} \Big ),
    \label{tl.X1.eq}
\end{align}
whereas the classical $X_2$-equation of motion yields (at zero external sources)
\begin{align}
    (\square + m^2)(X_1 + X_2) & = - (M^2 - m^2) X_2 - z \square X_2 + \frac{g_1 v}{ \Lambda^2} (D^\mu \phi)^\dagger D_\mu \phi + 
    \frac{g_2 v}{ \Lambda^2} F_{\mu\nu}^2 + \frac{g_3 v^3}{2  \Lambda^2} X_2^2.
    \label{tl.X2.eq}
\end{align}
By inserting Eqs.~(\ref{tl.X1.eq}) and~(\ref{tl.X2.eq}) into the solutions
of the $X_{1,2}$-equations (\ref{X12.sols}) we obtain the explicit form of the mapping for the HK model:
\begin{subequations}
\begin{align}
 \tbarc \rightarrow &
 - \frac{(M^2 - m^2)}{v^2} \Big ( \phi^\dagger \phi - \frac{v^2}{2} \Big )  - \frac{z}{v^2} \square \Big ( \phi^\dagger \phi - \frac{v^2}{2} \Big ) + \frac{g_1}{ \Lambda^2} (D^\mu \phi)^\dagger D_\mu \phi + 
    \frac{g_2}{ \Lambda^2} F_{\mu\nu}^2\nonumber \\
    & + \frac{g_3}{2  \Lambda^2} \Big ( \phi^\dagger \phi - \frac{v^2}{2} \Big )^2,
    \label{mapping1}\\
    \tT \rightarrow &\frac{g_1 }{\Lambda^2} \Big ( \phi^\dagger \phi - \frac{v^2}{2} \Big );\qquad 
    \tU \rightarrow \frac{g_2}{\Lambda^2}  \Big ( \phi^\dagger \phi - \frac{v^2}{2} \Big );\qquad
    \tR \rightarrow  \frac{g_3 v^2}{2 \Lambda^2} \Big ( \phi^\dagger \phi - \frac{v^2}{2} \Big ). 
\label{mapping2}
\end{align}
\label{mapping}
\end{subequations}

\section{Dimension six operators coefficients}\label{sec.dim6.coeffs}

For computing the UV coefficients of dimension 6 gauge-invariant operators in the target theory, we need to consider, see Appendix~\ref{app:g.inv.ops}:
\begin{enumerate}
    \item Operators which only depend on the external sources and contribute
    to dimension 6 operators in the target theory again due to the mapping in Eq.~(\ref{mapping}). They
    are listed in Eq.~(\ref{pure-ext.srcs}), and their UV coefficients denoted by~$\vartheta_i$'s;
    \item Mixed field-external sources gauge-invariant operators contributing to dimension 6 operators in the target theory under the mapping in Eq.~(\ref{mapping}); these
    are listed in Eq.~(\ref{mixed.field-ext.srcs}), and their UV coefficients by $\theta_i$'s;
    \item Dimension 6 field-dependent gauge-invariant operators that do not involve external sources; these are listed in Eq.~(\ref{ops.only.fields}) and their UV coefficients denoted by $\lambda_i$'s.
\end{enumerate}
Clearly, all the associated UV coefficients $\lambda_i$, $\theta_i$ and $\vartheta_i$ will be $\xi$-independent. In order to fix them, we need to evaluate a certain number of Feynman amplitudes and derive the projections of these operators on the relevant 1-PI Green's functions. However, and as already noticed, UV divergences of the latter cannot be parameterized in terms of the $\lambda_i$'s, $\theta_i$'s and $\vartheta_i$'s coefficients alone, since one needs to take into account contributions from GFRs. Indeed, the latter prove essential in order to ensure gauge independence of the UV coefficients of gauge invariant operators, as we will soon explicitly show.

\section{Generalized Field Redefinitions}\label{sec.gfrs}

The first and most difficult step for carrying out the off-shell renormalization program is to work out the GFRs controlled by ${\cal S}_0(\overline{Y}^{(1)})$. One needs to take them into account appropriately, otherwise the renormalization of gauge invariant operators is affected by spurious contributions arising from the incorrect subtraction of UV divergences to be removed by GFRs. In particular GFRs play a crucial role in ensuring the gauge independence of the UV coefficients of gauge invariant operators, as we will explicitly show.

In the Algebraic Renormalization approach we adopt, GFRs can be written in terms of two classes of invariants as
\begin{align}
    {\cal S}_0  \int \d \Big [ P(\Phi;\zeta) (\sigma^* \sigma + \chi^* \chi) + 
    Q(\Phi;\zeta) (\sigma^* (\sigma+v) + \chi^* \chi) \Big ],
    \label{af.1}
\end{align}
with $P$ and $Q$ some local functionals\footnote{We remind the reader that in EFTs field redefinitions are, in general, non-linear in the quantized fields.} depending on the fields  (collectively denoted by $\Phi$) and the external sources (collectively denoted by $\zeta)$ and ${\cal S}_0$ the linearized ST operator in Eq.~(\ref{S0}). For convenience, we refer to these terms as $P$- and $Q$-invariants. 

In order to get a better insight on the parameterization in Eq.~(\ref{af.1}) let us first consider the case where $P$ and $Q$ are constant. Since one has that
\begin{align}
    {\cal S}_0 \int \d \,  (\sigma^* \sigma + \chi^* \chi) =
    \int \d \,
    \Big [ \sigma 
    \frac{\delta \G^{(0)}}{\delta \sigma}
    + \chi
    \frac{\delta \G^{(0)}}{\delta \chi} +
    \sigma^* e \chi \omega -
    \chi^* e (v+\sigma) \omega\Big ] \supset
    - \int \d e v \chi^*\omega, 
\end{align}
the $P$-invariant is fixed in this case by the amplitude $\overline{\G}^{(1)}_{\omega\chi^*}$. Similarly, if $P$ depends on the fields and the gauge invariant sources $\bar c^*, R, T_1, U$, the $P$-invariant can be fixed by looking at antifield-dependent 1-PI amplitudes. Indeed, since the antighost equation~(\ref{antigh.eq}) entails that the dependence on the antighost at loops higher than one only happens via the combination $\widetilde{\chi}^*$ in Eq.~(\ref{chistar.redef}), we do not need to consider antighost amplitudes and antifield-dependent ones are sufficient.

The $Q$-invariant is trickier. Let us first notice that it does not project on $\chi^*,\sigma^*$ antifield-dependent monomials:
\begin{align}
 {\cal S}_0 & \int \d \,  (\sigma^* (\sigma + v) + \chi^* \chi)  =  \nonumber \\
    & \int \d \,
    \Big [ (\sigma + v) 
    \frac{\delta \G^{(0)}}{\delta \sigma}
    + \chi
    \frac{\delta \G^{(0)}}{\delta \chi} +
    \sigma^* e \chi \omega -
    \chi^* e (v+\sigma) \omega\Big ] 
    \supset\int \d \, v^2 \bar c^*
    - \int \d \, v m^2\sigma. 
    \label{Q-inv}
\end{align}
However, \1eq{Q-inv} clearly shows that it yields a contribution to $\bar c^*$ (and the $\sigma$-tadpole).

To understand the $Q$-invariant role in the renormalization of the theory, we remark that it depends only on the combination $\phi_0$; therefore it is useful to rewrite the counting operator in terms of $\phi, \phi^\dagger$, {\it i.e.}, 
\begin{align}
    {\cal S}_0 
    \int \d \,  (\sigma^* (\sigma + v) + \chi^* \chi)  = 
    {\cal S}_0 
    \int \d \,
    \Big ( 
    \phi \frac{\delta \G^{(0)}}{\delta \phi}
    +
    \phi^\dagger \frac{\delta \G^{(0)}}{\delta \phi^\dagger}    
    \Big ).
\end{align}
Next, observe that we are only interested in the case when the right-hand side (r.h.s.) is evaluated at $X_{1,2}=0$\footnote{$X_{1,2}$-amplitudes being fixed in a purely algebraic way by Eq.~(\ref{nth.X12.eqs})}; an explicit computation shows that the r.h.s. is indeed gauge-invariant (remember that we need to use the antifield $\widetilde{\chi}^*$, as a consequence of the antighost equation):
\begin{align}
     {\cal S}_0 
    \int \d \,
    \Big ( 
    \phi \frac{\delta \G^{(0)}}{\delta \phi}
    +
    \phi^\dagger \frac{\delta \G^{(0)}}{\delta \phi^\dagger}    
    \Big )  
    & = \int d^4 x \, \Big [ -2 \phi^\dagger D^2 \phi -
    \frac{2 m^2}{v^2} \Big ( \phi^\dagger \phi - \frac{v^2}{2} \Big ) \phi^\dagger \phi \nonumber \\
    & -
    \partial^\mu T_1 ( \phi^\dagger D_\mu \phi + \mathrm{h.c.}) - T_1 (\phi^\dagger D^2 \phi + \mathrm{h.c.}) +
    2 \bar c^* \phi^\dagger \phi \Big ].
    \label{2nd.cohom.triv.inv}
\end{align}
Notice in particular that the dependence on
$\sigma^*, \chi^*$ has disappeared; as a consequence this invariant  contains a combination of gauge-invariant operators that vanish on-shell. Let us now consider what happens in the power-counting renormalizable case ($T_1=0$ and $z=g_i=0$). Imposing the mapping in Eq.~(\ref{mapping}) on the r.h.s. of Eq.~(\ref{2nd.cohom.triv.inv}) we obtain\footnote{Observe that as announced the $m^2$-dependence has disappeared.}:
\begin{align}
     {\cal S}_0 
    \int \d \,
    \Big ( 
    \phi \frac{\delta \G^{(0)}}{\delta \phi}
    +
    \phi^\dagger \frac{\delta \G^{(0)}}{\delta \phi^\dagger}    
    \Big ) = 
    - \int \d \Big [ 
    2 \phi^\dagger D^2 \phi + \frac{2 M^2}{v^2} \Big ( \phi^\dagger \phi - \frac{v^2}{2} \Big )^2
    - M^2 \Big ( \phi^\dagger \phi - \frac{v^2}{2} \Big )
    \Big ].
    \label{field.redef.comb}
\end{align}
On the other hand, the gauge-invariant operators of the renormalizable Abelian HK model with dimension $\leq 4$ are
\begin{align}
    &\int \d F_{\mu\nu}^2;&
    &\int \d (D^\mu \phi)^\dagger D_\mu \phi;&
    &\int \d \Big ( \phi^\dagger \phi - \frac{v^2}{2} \Big );&
    &\int \d \Big ( \phi^\dagger \phi - \frac{v^2}{2} \Big )^2,
\end{align}
whereas the number of physical parameters is $3$, which are usually chosen to be: the gauge coupling $e$ associated with the coefficient of the field strength squared; the mass of the vector meson $M_A$, which is related to the renormalization of the vev via the tadpole invariant; and, finally, the mass of the physical scalar $M$, which appears with the quartic potential invariant. The scalar kinetic covariant term is related instead to the wave function renormalization of the two-point Higgs field and as such cannot have physical effects. If we denote by $Z^{1/2}$ the coefficient of the corresponding invariant (\ref{field.redef.comb}), the combination in the r.h.s. of that equation is exactly the one related to the wave function renormalization $\phi \rightarrow (1 + Z^{1/2}) \phi$.

Motivated by these remarks, we choose to express all $Q$-invariants in the $X$-theory of the form
\begin{align}
    \int \d \, Q(\Phi;\zeta) \phi^\dagger D^2 \phi, 
\end{align}
with $Q(\Phi;\zeta)$ gauge-invariant, as a linear combination of gauge invariant operators and cohomologically trivial invariants of the form
\begin{align}
	{\cal S}_0  \int \d  Q(\Phi;\zeta)(\sigma^* (\sigma + v) + \chi^* \chi).
\end{align}
This provides a consistent definition of the independent gauge invariant operators  generalizing the corresponding set of independent physical parameters discussed in the power-counting renormalizable case.

We also notice that in the Landau gauge ($\xi=0$) ghosts are free and the theory enjoys an exact global invariance 
\begin{align}
	\delta \phi &= i e \alpha \phi;&
	\delta\phi^\dagger &= - i e \alpha \phi^\dagger
\end{align} 
with $\alpha$ a constant parameter. As a consequence of this rigid $U$(1) invariance the only allowed cohomologically trivial invariants in the Landau gauge are those of the $Q$-type; $P$-invariants do not arise. We will verify this property in the explicit computations that follow. On the other hand, notice  that in a general gauge, $Q$ need not be gauge-invariant and both $P$ and $Q$-type invariants are required, due to the fact that  the vev renormalizes differently than the fields, as is well known in the literature~\cite{Sperling:2013eva}.

We now list the monomials in the expansion of $P,Q$ contributing to the projections needed to fix the coefficients of the dimension 6 operators in Eqs.~(\ref{pure-ext.srcs}),~(\ref{mixed.field-ext.srcs}) and~(\ref{ops.only.fields}).
Using the notation
\begin{align}
{\cal Z}_{1} &\equiv (\sigma^* \sigma  + \chi^* \chi);& 
{\cal Z}_{2} &\equiv (\sigma^* (\sigma + v) + \chi^* \chi),
\end{align}
we obtain
\begin{align}
    \overline{Y}^{(1)} = {\cal S}_0 \int \d \Big [ 
    & \Big ( \rc{0} + \rc{1} \sigma + \rc{2} \sigma^2 + \rc{3} \chi^2 + \rc{0T} T_1 \Big ) {\cal Z}_{1} \nonumber \\
    & + \Big ( \trc{0} + \trc{1} \sigma + \trc{2} \sigma^2 
    + \trc{3} \chi^2 +
    \trc{4} \sigma \chi^2 \nonumber \\
    & \quad + \trc{0T} T_1  +  \trc{0TT} T_1^2 + \trc{1T} T_1 \sigma + 
    \trc{3T} T_1 \chi^2 
    \Big ) {\cal Z}_{2}
    \Big ].
    \label{cohom.triv.ol}
\end{align}
The different coefficients can be then evaluated by projection onto the relevant Feynman amplitudes; their values are then
\begin{subequations}
\begin{align}
    \rc{0} &= \frac{( 1 - \delta_{\xi;0} )}{8 \pi^2 v^2}
    \frac{M_A^2}{1+z} 
    \frac{1}{\epsilon};&
    \rc{1} &= -\frac{( 1 - \delta_{\xi;0} )}{4 \pi^2 v^3}
    \frac{z M_A^2}{(1+z)^2},&  \\
    \rc{2} &= \frac{( 1 - \delta_{\xi;0} )}{8 \pi^2 v^4}
    \frac{z (3z - 1) M_A^2}{(1+z)^3} 
    \frac{1}{\epsilon}; &
    \rc{3} &= - \frac{( 1 - \delta_{\xi;0} )}{8 \pi^2 v^4}
    \frac{z  M_A^2}{(1+z)^2} 
    \frac{1}{\epsilon},& \\
    \rc{0T} &=  - \frac{( 1 - \delta_{\xi;0} )}{8 \pi^2 v^2}
    \frac{M_A^2}{(1+z)^2} 
    \frac{1}{\epsilon};& 
    \trc{0} &= \frac{(1 - \delta_{\xi;1})}{16 \pi^2 v^2} M_A^2,\\
    \trc{1} &= -\frac{(1 - \delta_{\xi;1})}{8 \pi^2 v^3} \frac{z M_A^2}{1+z};& 
    \trc{2} &= \frac{(1 - \delta_{\xi;1})}{8 \pi^2 v^4} \frac{z(z-1) M_A^2}{(1+z)^2}, \\
    \trc{3} &= \frac{(-1)^{\delta_{\xi;0}}}{16 \pi^2 v^4} \frac{z M_A^2}{1+z};&
    \trc{4} &= -(-1)^{\delta_{\xi;0}}
      \frac{ z [ 3z + (-1)^ {\delta_{\xi;0}}  ]
      }{16 \pi^2 v^5 (1+z)^2}\frac{M_A^2}{\epsilon},\\
    \trc{0T} &= - \frac{( 1 - \delta_{\xi;1} )}{8 \pi^2 v^2} \frac{M_A^2}{\epsilon};&
    \trc{0TT} &= \frac{( 1 - \delta_{\xi;1} )}{8 \pi^2 v^2} \frac{M_A^2}{\epsilon},\\
    \trc{1T} &= \frac{( 1 - \delta_{\xi;0} )}{8 \pi^2 v^3} \frac{z ( 2 + z) M_A^2}{(1+z)^2} \frac{1}{\epsilon};& 
    \trc{3T} &= 0.
\end{align}
\label{cohom.triv.ol.values}
\end{subequations} 

Notice  that in Landau gauge $\overline{Y}^{(1)}$ reduces to
\begin{align}
  \left . \overline{Y}^{(1)}  \right |_{\xi= 0 } = 
  {\cal S}_0 & \int \d \frac{M_A^2}{32 \pi^2 v^2}
  \frac{1}{\epsilon}
  \Big [  2  - 4 T_1 + 4 T_1^2 
  \nonumber \\
  & - \frac{4}{v^2} \frac{z}{1+z}
  \Big ( \phi^\dagger \phi - \frac{v^2}{2} \Big ) + 
  \frac{2}{v^4}
  \frac{z ( 3z -1)}{(1+z)^2}
  \Big ( \phi^\dagger \phi - \frac{v^2}{2} \Big )^2 
 \Big ] {\cal Z}_{2},
 \label{Y1.landau}
\end{align}
{\it i.e.}, the polynomial $Q$ is gauge-invariant, as expected; moreover, as anticipated, all $\rho$'s coefficients vanish in this gauge.

\subsection{GFRs in the target theory}
\label{sec:GFRphi}

It is instructive to obtain the explicit form of the GFRs in the  target theory at linear order in the higher dimensional couplings. For that purpose we need to apply the mapping in Eq.~(\ref{mapping}) to $\overline{Y}^{(1)}$ retaining only the terms linear in the $g_i$'s and $z$.

We remark that the coefficients in Eq.~(\ref{cohom.triv.ol.values}) only depend on $z$. Moreover, the image of the source $T_1$ under the mapping is proportional to $g_1$ and hence from the $T_1$ sector we receive contributions at the linearized level only from amplitudes linear in $T_1$, whose coefficients need to be evaluated at $z=0$. By taking these observations into account, one easily sees that the GFRs in the target theory at linear order in the $g_i$'s and $z$ couplings take the following form:
\begin{align}
     \begin{pmatrix}\phi_0' \cr \chi' \cr\end{pmatrix} &= 
      \Bigg \{
      1 + \frac{M_A^2}{16 \pi^2 v^2}
      \Bigg [
      2 (1 - z) (1 - \delta_{\xi;0}) +
      (1 - \delta_{\xi;1})
      -2 \Big [ 
       (1 - \delta_{\xi;0}) \Big ( \frac{g_1 v}{\Lambda^2} +  \frac{2z}{v} \Big )
       \nonumber \\
       &+
       (1 - \delta_{\xi;1})
       \Big ( \frac{g_1 v}{\Lambda^2} + \frac{z}{v}  \Big )
      \Big ] \sigma
      - 
      \Big ( \frac{2z}{v^2} + \frac{g_1}{\Lambda^2} \Big )
       \sigma^2 
       -
       \Big ( \frac{z}{v^2} + \frac{g_1}{\Lambda^2} \Big )
       \chi^2
      -\frac{z}{v^3} \sigma \chi^2 + \cdots
      \Bigg ]
      \Bigg \} \frac{1}{\epsilon}  
      \begin{pmatrix}\sigma \cr \chi \cr\end{pmatrix} \nonumber \\
      & + \Bigg \{ 1 + \frac{M_A^2}{16 \pi^2 v^2} 
      \Bigg [ 
      1 - \delta_{\xi;1} 
      - 2 (1 - \delta_{\xi;1}  )  \Big ( 
      \frac{z}{v} + \frac{g_1 v}{\Lambda^2} 
      \Big )  \sigma - \frac{z}{v^3} \sigma \chi^2- (1 - \delta_{\xi;1}  ) \Big ( \frac{2z}{v^2} + \frac{g_1}{\Lambda^2} \Big ) \sigma^2\nonumber \\
      & - \Big [ (-1)^{1+\delta_{\xi;0}} \frac{z}{v^2} + 
      (1 - \delta_{\xi;1}) \frac{g_1}{\Lambda^2} \Big ]
      \chi^2 + \cdots
      \Bigg ]
      \Bigg \} 
      \frac{1}{\epsilon}  
      \begin{pmatrix}v \cr 0 \cr\end{pmatrix},
    \label{GFR.phi}
    \end{align}
where the dots denote higher dimensional contributions that are not relevant in the one loop renormalization of the dimension 6 operators under consideration.
Notice also that the  contribution proportional to the constant spinor
$(v,0)^\mathrm{T}$ is associated with the $Q$-type invariants.

From Eq.(\ref{GFR.phi}) we see  that the GFRs are non-multiplicative already at one loop and in the linearized approximation.

\section{Renormalization of Gauge Invariant Operators}\label{sec.ren}

Once the cohomologically trivial sector has been fixed as in Eq.(\ref{cohom.triv.ol}) and (\ref{cohom.triv.ol.values}) we can proceed to project on the one-loop amplitudes required to determine the coefficients of the invariants (\ref{pure-ext.srcs}), (\ref{mixed.field-ext.srcs}) and (\ref{ops.only.fields}). As the methodology is illustrated in detail in Ref.~\cite{Binosi:2019nwz}, we report here only the results, which have been explicitly evaluated in both Landau and Feynman gauge and found to coincide as required.

\subsection{Pure external sources invariants}
The non zero $\vartheta_i$ coefficients are 
\begingroup
\allowdisplaybreaks
\begin{subequations}
\begin{align}
	\cfps{1} &=  -\frac{1}{16 \pi^2} \frac{M^2 + (1+z)^2 M_A^2 }{(1+z)^2} \frac{1}{\epsilon};&
	\cfps{2} &= \frac{1}{16 \pi^2} \frac{-M^4 + 3 (1+z)^3 M_A^2 }{(1+z)^3} \frac{1}{\epsilon},\\
	\cfps{3} &= \frac{3 M_A^4}{4\pi^2}\frac{1}{\epsilon};&
	\cfps{4} &= - \frac{1}{8 \pi^2} \frac{M^2}{(1+z)^2}\frac{1}{\epsilon},\\
	\cfps{5} &= \frac{1}{16 \pi^2}\frac{2 + 2z + z^2}{(1+z)^2}\frac{1}{\epsilon};&
	\cfps{6} &= \frac{3}{16 \pi^2} \frac{M^4 + (1 + z)^4 M_A^4 }{(1+z)^4}\frac{1}{\epsilon},\\
	\cfps{7} &= \frac{9 M_A^4}{\pi^2}\frac{1}{\epsilon};& 
	\cfps{8} &= \frac{1}{4 \pi^2 (1+z)^2}\frac{1}{\epsilon},\\
	\cfps{10} &= \frac{3}{32 \pi^2} \frac{M^2 + (1+z)^3 M_A^2}{(1+z)^3} \frac{1}{\epsilon};&
	\cfps{11} &= \frac{3 M_A^2}{\pi^2}\frac{1}{\epsilon},\\
	\cfps{13} &= \frac{1}{8 \pi^2} \Big [ M_A^2 + \frac{M^2}{(1+z)^3} \Big ] \frac{1}{\epsilon};&
	\cfps{15} &= \frac{1}{8 \pi^2 (1+z)^2}\frac{1}{\epsilon},\\
	\cfps{16} &= \frac{3 M_A^4}{2 \pi^2}\frac{1}{\epsilon};&
	\cfps{17} &= \frac{1}{4 \pi^2}\frac{M_A^2}{(1+z)^3}\frac{1}{\epsilon},\\
	\cfps{19} &= \frac{1}{32 \pi^2}
	\frac{2 + 2z + z^2}{(1+z)^2}\frac{1}{\epsilon};&
	\cfps{22} &= \frac{3 M_A^2}{8 \pi^2} \frac{1}{\epsilon},\\
	\cfps{23} &= \frac{1}{16 \pi^2 (1+z)^2} \frac{1}{\epsilon};& 
	\cfps{26} &= -\frac{3}{4 \pi^2} \frac{M_A^4}{(1+z)^5}\frac{1}{\epsilon},\\
	\cfps{27} &= \frac{144 M_A^4}{\pi^2} \frac{1}{\epsilon};&
	\cfps{29} &= -\frac{1}{8 \pi^2} \frac{2 + 3z + 3z^2 + z^3}{(1+z)^3} \frac{1}{\epsilon},\\
	\cfps{32} &= - \frac{1}{8 \pi^2}\frac{3 M^2 + 2 (1+z)^4 M_A^2}{(1+z)^4}\frac{1}{\epsilon};&
	\cfps{36} &= -\frac{1}{4 \pi^2}\frac{1}{(1+z)^3}\frac{1}{\epsilon},\\
	\cfps{38} &= \frac{3 M_A^4}{2\pi^2}\frac{1}{\epsilon};&
	\cfps{39} &= -\frac{3 M^2}{4 \pi^2 (1+z)^4}\frac{1}{\epsilon},\\ 
	\cfps{40} &= \frac{18 M_A^4}{\pi^2}\frac{1}{\epsilon};&
	\cfps{41} &= -\frac{1}{2\pi^2} \frac{1}{(1+z)^3}\frac{1}{\epsilon}.
	\end{align}
\label{cfps.values}
\end{subequations}
\endgroup

\subsection{Mixed field-external sources invariants}
The non zero $\cfxt{i}$ coefficients are
\begingroup
\allowdisplaybreaks
\begin{subequations}
	\begin{align}
	\cfxt{1} &= -\frac{1}{16 \pi^2 v^2}\frac{1}{(1+z)^3}\Big [2 (1-z) M^2 + 2 (1+z)^2 M_A^2 + (2 + 4 z + 3 z^2 + z^3) m^2
    \Big ] \frac{1}{\epsilon},\\
    \cfxt{2} &= \frac{1}{8 \pi^2 v^2}
    \frac{1}{(1+z)^4}
    \Big \{
    (z-2) M^4 + 6 (1+z)^4 M_A^4 - (1+z) 
    \Big [ M^2 + (1+z)^3 M_A^2 \Big ] m^2
    \Big \} \frac{1}{\epsilon},\\
    \cfxt{3} & = \frac{3M_A^4 }{\pi^2 v^2}\frac{1}{\epsilon},\\
    \cfxt{4} &= - \frac{1}{8 \pi^2 v^2}\frac1{(1+z)^3} \Big[(1+z) m^2  + 4 M^2\Big]\frac{1}{\epsilon},\\
    \cfxt{5} &= - \frac{1}{32 \pi^2 v^2}\frac1{(1+z)^2}
    \Big [ 
    -4z (1+z)+ 4(1+z)
    \frac{ g_1 v^2}{\Lambda^2}
    + (2 + z) \frac{g_1^2 v^4}{\Lambda^4}
    \Big ] \frac{1}{\epsilon},\\
    \cfxt{6} &= -\frac{1}{32 \pi^2 v^2 }\frac1{(1+z)^3} 
    \Big \{ 
    2 ( 2 + 4z + 3z^2 + z^3) m^2 -
    (1+z) \Big [ 
    -4 + 8 z^2 +\nonumber \\
    &- \Big ( 8  + \frac{12 g_1 v^2}{\Lambda^2} \Big ) z + 
    \frac{3 g_1^2 v^4}{\Lambda^4} 
    \Big ] M_A^2
    + \Big [ 4 + \frac{g_1^ 2 v^ 4}{\Lambda^4} ( 3 + z) \Big ] M^ 2,
    \\
    \cfxt{7} &= -\frac{3 M_A^2}{8 \pi^2 v^2}\frac1{(1+z)}
    \Big [ 
    -4 z + \frac{g_1 v^2}{\Lambda^4} 
    \Big ( g_1 v^2 + 4 \Lambda^2 \Big )
    \Big ] \frac{1}{\epsilon},\\
    \cfxt{8} &= -
    \frac{g_1^2 v^2}{16 \pi^2\Lambda^4}\frac{1}{ (1+z)^2}\frac{1}{\epsilon},\\
    \cfxt{9} &= -
    \frac{g_1}{16 \pi^2 \Lambda^2}\frac{1}{(1+z)^2}\frac{1}{\epsilon},\\
    \cfxt{10} &= -\frac{1}{32 \pi^2 v^2}\frac1{(1+z)^3}
    \Big [ 
    -2 (1+z)^3 M_A^2  
    + \Big ( 2 + \frac{4 g_1 v^2}{\Lambda^2} \Big ) M^2 +  
    (2 + 4z + 3z^2 + z^3) m^2
    \Big ]\frac{1}{\epsilon},\\
    \cfxt{11} &= \frac{3 M_A^2}{4\pi^2 v^2} \frac{1}{\epsilon}, \\
    \cfxt{12} &= - \frac{g_1}{8 \pi^2 \Lambda^2 }\frac1{(1+z)^2}\frac{1}{\epsilon},\\
    \cfxt{13} &= \frac{1}{8 \pi^2 v^4}\frac{z}{(1+z)^4}
    \Big [ (1+z)^2 (5+z) M_A^2 + 
    4 (2-z) M^2 + 4 (1+z) m^2
    \Big ] \frac{1}{\epsilon},\\ 
    \cfxt{14} &= -\frac{1}{8 \pi^2 v^4}
    \frac{1}{(1+z)^5}
    \Big \{ 
    (1+z)^2 (2 + 3z + 3z^2 + z^3) m^4 + 
    4 (1+z) m^2 \Big [ 
    (1-2z) M^2 \nonumber \\
    &+ (1+z)^2 M_A^2
    \Big ] + 4 \Big [ 
     -3 (1+z)^5 M_A^4 + (1-4 z + z^2) M^4
     \Big ]
    \Big \} 
    \frac{1}{\epsilon}, \\ 
    \cfxt{15} &=  \frac{6 M_A^4}{\pi^2 v^4}\frac{1}{\epsilon},\\ 
    \cfxt{16} &= \frac{1}{2\pi^2 v^4}\frac1{(1+z)^4}
    \Big [ 
    2 ( 2z - 1) M^2 + (z^2 - 1) m^2 
    \Big ] \frac{1}{\epsilon},\\
    \cfxt{17} &= \frac{g_2^2 v^2}{8 \pi^2  \Lambda^4}\frac1{(1+z)^2} \frac{1}{\epsilon},\\
    \cfxt{18} &= - \frac{1}{256 \pi^2 v^2 \Lambda^4}\frac1{(1+z)^3}
    \Big \{ 
    -64 g_2^2 v^4 M^2 + 
    (1+z) \Big [ (2+z) g_1^2 v^4 +    4 g_1 v^2 z ( 2 g_2 v^2 + \Lambda^2)  \nonumber \\
    &+ 4z ( 8 g_2^2 v^4 + 4 g_2 v^2 \Lambda^2 - (1+z) \Lambda^4 \Big ] M_A^2 \Big \} 
    \frac{1}{\epsilon}, \\
    \cfxt{19} &= - \frac{g_2^2 v^2}{2 \pi^2  \Lambda^4}\frac1{(1+z)^2}
    \Big [ M^2 + 2(1+z) M_A^2 \Big ] \frac{1}{\epsilon},\\
    \cfxt{20} &= \frac{g_2^2 v^2}{4 \pi^2 \Lambda^4}\frac1{(1+z)^2 }\frac{1}{\epsilon},\\
    \cfxt{21} &= - \frac{1}{4 \pi^2 v^2}\frac{z}{(1+z)^3}
    \frac{1}{\epsilon},\\
    \cfxt{22} &= \frac{1}{8 \pi^2 v^2}\frac1{(1+z)^5}
    \Big[6 ( 1 - z) M^4 + 
    6 ( 1 + z)^5 M_A^4 + (1+z) (3 M^2 + 2 (1+z)^4 M_A^2) m^2 \Big]
    \frac{1}{\epsilon}, \\
	\cfxt{23} &= \frac{36 M_A^4}{\pi^2 v^2}\frac{1}{\epsilon}, \\
	\cfxt{24} &= \frac{1}{\pi^2 v^2}\frac1{(1+z)^3}\frac{1}{\epsilon}, \\
	\cfxt{25} &= \frac{1}{8 \pi^2 v^2}\frac1{(1+z)^4}\Big[
    (2 - 4z) M^2 + 2  (1+z)^2 M_A^2 + 
    (2 + 5z + 6 z^2 + 4 z^3 + z^4) m^2\Big]
    \frac{1}{\epsilon}, \\
    \cfxt{27} & = \frac{1}{4 \pi^2 v^2}\frac{ (1-z)}{(1+z)^3} \frac{1}{\epsilon},\\ 
    \cfxt{28} &= \frac{6 M_A^4}{\pi^2 v^2}\frac{1}{\epsilon},\\
    \cfxt{29} &= \frac{1}{4 \pi^2 v^2}\frac1{(1+z)^4}
    \Big[2 ( 2 - z) M^2 + (1+z) m^2 \Big] 
    \frac{1}{\epsilon}. 
    \end{align}
    \label{cfxt.values}
\end{subequations}
\endgroup
 
\subsection{Gauge invariants depending only on the fields}
The non zero $\lambda_i$ coefficients are
\begingroup
\allowdisplaybreaks
\begin{subequations}
	\begin{align}
	\cfgi{1} &= \frac{1}{16 \pi^2 v^2}\frac1{(1+z)^3}\Big\{
	(1+z) \Big [M^2 + (1+z)^2 M_A^2  \Big ] m^2
	+ 2 \Big [ M^4 + 3 (1+z)^3 M_A^4 \Big ]\Big\}\frac{1}{\epsilon},\\
	\cfgi{2} &= \frac{1}{32 \pi^2 v^4}\frac{1}{(1+z)^4} \Big \{ 4 ( 1 - 2z) M^4 + 4 m^2 M_A^2 (1+z)^3 + 12 (1+z)^4 M_A^4 \nonumber \\ 
	& + 4 m^2 M^2 (1-z^2) +  (1+z)^2 (2 + 2z + z^2)m^4 \Big \}  \frac{1}{\epsilon},\\
	\cfgi{3} &= -\frac{1}{16 \pi^2 v^6}\frac{z}{(1+z)^5}\Big \{ 8 (1-z) M^4 + 2 (1+z)^2 m^4 + (1+z) \Big [ 4  (2-z) M^2 \nonumber \\
	&+ (1+z)^2(5+z) M_A^2\Big ] m^2\Big \} \frac{1}{\epsilon},\\
	\cfgi{4} & = -\frac{1}{32 \pi^2 v^2}\frac1{(1+z)^2}\Big \{ (1+z) \Big [ 16 + 4 z + \frac{3 g_1^2 v^4}{\Lambda^4} + 12 \frac{ g_1 v^2}{\Lambda^2} \Big ] M_A^2+\frac{g_1 v^2}{\Lambda^2} \Big ( 4 - \frac{g_1 v^2}{\Lambda^2} \Big )M^2 \Big \} \frac{1}{\epsilon},	\\
	\cfgi{5} & = \frac{g_1^2 v^2}{192 \pi^2\Lambda^4 }\frac{1}{(1+z) }\frac{1}{\epsilon},\\
	\cfgi{6} &= \frac{1}{64 \pi^2 v^4}\frac{1}{(1+z)^3}\Big \{\Big [4z + 4 ( 1 - 3z) \frac{g_1 v^2}{\Lambda^2}+(1+z) \frac{g_1^2 v^4}{\Lambda^4}\Big ] M^2 \nonumber \\
	&+(1+z)^2\Big ( 4 z - 12 \frac{g_1 v^2}{\Lambda^2} - 3 \frac{g_1^2 v^4}{\Lambda^4} \Big ) M_A^2+ 4 ( 1 + z) \frac{g_1 v^2}{\Lambda^2} m^2\Big \} \frac{1}{\epsilon},\\
	\cfgi{7} &= \frac{1}{32 \pi^2 v^4}\frac{1}{(1+z)^3}\Big \{ \Big [ 4z - 4 (1+z) \frac{g_1 v^2}{\Lambda^2}+ (5+z) \frac{g_1^2 v^4}{\Lambda^4} \Big ]M^2 \\ 
	&+ 3 (1+z) \Big [ 4z (3+z) - 4 (3+z) \frac{g_1 v^2}{\Lambda^2} - (5+ 3z) \frac{g_1^2 v^4}{\Lambda^4}\Big ] M_A^2 \nonumber \\
	& + (1+z) \Big [ - 4 z(1+z) + 4 (1+z) \frac{g_1 v^2}{\Lambda^2}+ (2+z) \frac{g_1^2 v^4}{\Lambda^4} \Big ] m^2\Big \} \frac{1}{\epsilon},\\
	\cfgi{8} &= -\frac{1}{192 \pi^2 v^2}\frac{1}{(1+z)^2} \Big \{ 48 \frac{g_2^2 v^4}{\Lambda^4} M^2 \nonumber \\
	&+ (1+z) \Big [ 2 + 48 \Big ( \frac{g_2 v^2}{\Lambda^2}+ \frac{g_2^2 v^4}{\Lambda^4}\Big ) +2 g_1 \frac{v^2}{\Lambda^2} + 24 g_1 g_2 \frac{v^4}{\Lambda^4}  + g_1^2 \frac{v^4}{\Lambda^4}\Big ] M_A^2 \Big \} \frac{1}{\epsilon}, \\
	\cfgi{9} & = - \frac{g_2^2 v^2}{6 \pi^2\Lambda^4 }\frac{1}{(1+z)}\frac{1}{\epsilon},\\
	\cfgi{10} & = \frac{1}{128 \pi^2 v^2 \Lambda^2}\frac{1}{(1+z)^3}\Big \{ - 64 \frac{g_2^2 v^2}{\Lambda^2} M^2 + (1+z) \Big [ g_1^3 \frac{v^4}{\Lambda^4} - 4 \frac{g_1^2 v^2}{\Lambda^2}\nonumber \\
	&- 8 (2+z) g_1 g_2 \frac{v^2}{\Lambda^2}- 16 (2 + z)\Big ( 2 g_2^2\frac{v^2}{\Lambda^2} + g_2 \Big ) \Big ] M_A^2 - 16 (1+z) \frac{g_2^2 v^2}{\Lambda^2} m^2\Big \} \frac{1}{\epsilon}.
	\end{align}
	\label{cfgi.values}
\end{subequations}
\endgroup

\section{Mapping to the target theory}

The UV coefficients in the target theory $\cfgit{i}$ can be obtained by: applying the mapping in Eq.~(\ref{mapping}) to the invariants in Eqs.~(\ref{pure-ext.srcs}) and~(\ref{mixed.field-ext.srcs}); combining the projections with the operators in~(\ref{ops.only.fields}); and, finally, using the results~(\ref{cfps.values}), (\ref{cfxt.values}) and (\ref{cfgi.values}). Notice that for these coefficients all $m^2$-dependent contributions must cancel out; we have checked this explicitly.

The coefficients so obtained represents the complete one-loop renormalizations of the corresponding operators; in particular, no linearized approximation in the higher dimensional couplings $g_i$'s has been made so far. However, as the resulting general expressions are rather lengthy, we report below the non zero coefficients $\cfgit{i}$ at linear order in the $g_i$ couplings:
\begingroup
\allowdisplaybreaks
\begin{subequations}
\begin{align}
	\cfgit{1} &\sim -\frac{1}{16 \pi^2 v^2}
    \Big [ 8 z M_A^4 + (M^4 - 3 M_A^4) \frac{g_1 v^2}{\Lambda^2}  - 12 M_A^4 \frac{g_2 v^2}{\Lambda^2}
    + M^2 g_3  \frac{v^4}{\Lambda^2} \Big ]
    \frac{1}{\epsilon},\\
    \cfgit{2} &\sim -\frac{1}{32 \pi^2 v^4}
    \Big [ 
    \Big ( 4 M^2 M_A^2 + 42 M^4 \Big ) z 
    + 4 \Big ( 3 M^4 + M^2 M_A^2 - 6 M_A^4 \Big )
    \frac{g_1 v^2}{\Lambda^2} \nonumber \\
    &
    - 96 \frac{g_2 v^2}{\Lambda^2} M_A^4
    + \Big ( 11 M^2 + M_A^2 \Big ) \frac{g_3 v^4}{\Lambda^2} 
    \Big ] \frac{1}{\epsilon},\\
    \cfgit{3} &\sim -\frac{1}{16 \pi^2 v^6}
    \Big [ 
    z M^2 ( 18 M^2 + 5 M_A^2 ) +
    2 ( 5 M^4 + 2 M^2 M_A^2 - 6 M_A^4) \frac{g_1 v^2}{\Lambda^2} \nonumber \\
    &
    - 48 M_A^4 \frac{g_2 v^2}{\Lambda^2} +
    ( 8 M^2 + M_A^2)
    \frac{g_3 v^4}{\Lambda^2}
    \Big ]\frac{1}{\epsilon},\\
    \cfgit{4} &\sim \frac{1}{16 \pi^2 v^2}
    \Big [ 6 z M_A^2 -  ( 3 M^2 + 7 M_A^2  ) \frac{g_1 v^2}{\Lambda^2} \Big ] \frac{1}{\epsilon},\\
    \cfgit{6} &\sim \frac{1}{16 \pi^2 v^4}
   \Big [ 
   z ( 5 M^2 + 3 M_A^2) - 2 M_A^2 \frac{g_1 v^2}{\Lambda^2} + 12 M_A^2 \frac{g_2 v^2}{\Lambda^2}
   \Big ] \frac{1}{\epsilon},\\
   \cfgit{7} &\sim  \frac{1}{8 \pi^2 v^4}  ( 4 M^2 + 11 M_A^2  ) \Big (z - \frac{g_1 v^2}{\Lambda^2} \Big ) \frac{1}{\epsilon},\\
   \cfgit{8} &\sim - \frac{1}{96 \pi^2 v^2}
   \Big [  6 \frac{g_2 v^2}{\Lambda^2} (M^2 + 5 M_A^2) + M_A^2 \Big (- z + \frac{g_1 v^2}{\Lambda^2} \Big ) \Big ] \frac{1}{\epsilon},\\
   \cfgit{10} &\sim -\frac{1}{8 \pi^2 v^4} \frac{g_2 v^2}{\Lambda^2}  (2 M^2 + M_A^2)\frac{1}{\epsilon}.
\end{align}
\label{tilde.lambdas}
\end{subequations}
\endgroup
We hasten to emphasize that  GFRs do contribute also at the linearized level, as has been discussed in detail in Section~\ref{sec:GFRphi}. Failure to take their contributions into account would lead to an erroneous determination of the coefficients in Eq.(\ref{tilde.lambdas}).

The $g_i$'s, $z$ contributions to the $\beta$ functions 
\begin{align}
    \beta_i = (4 \pi)^2 \frac{d}{d \log \mu} \tilde{\lambda}_i
\end{align}
can then be easily determined from Eq.~(\ref{tilde.lambdas}), leading to: 
\begingroup
\allowdisplaybreaks
\begin{subequations}
\begin{align}
	\beta_1 &\supset -\frac{1}{v^2}
    \Big [ 8 z M_A^4 + (M^4 - 3 M_A^4) \frac{g_1 v^2}{\Lambda^2}  - 12 M_A^4 \frac{g_2 v^2}{\Lambda^2}
    + M^2 g_3  \frac{v^4}{\Lambda^2} \Big ],\\
    \beta_2 &\supset -\frac{1}{v^4}
    \Big [ 
    \Big ( 4 M^2 M_A^2 + 42 M^4 \Big ) z 
    + 4 \Big ( 3 M^4 + M^2 M_A^2 - 6 M_A^4 \Big )
    \frac{g_1 v^2}{\Lambda^2} \nonumber \\
    &
    - 96 \frac{g_2 v^2}{\Lambda^2} M_A^4
    + \Big ( 11 M^2 + M_A^2 \Big ) \frac{g_3 v^4}{\Lambda^2} 
    \Big ] ,\\
    \beta_3 &\supset -\frac{3}{v^6}
    \Big [ 
    z M^2 ( 18 M^2 + 5 M_A^2 ) +
    2 ( 5 M^4 + 2 M^2 M_A^2 - 6 M_A^4) \frac{g_1 v^2}{\Lambda^2} \nonumber \\
    &
    - 48 M_A^4 \frac{g_2 v^2}{\Lambda^2} +
    ( 8 M^2 + M_A^2)
    \frac{g_3 v^4}{\Lambda^2}
    \Big ],\\
    \beta_4 &\supset \frac{1}{v^2}
    \Big [ 6 z M_A^2 -  ( 3 M^2 + 7 M_A^2  ) \frac{g_1 v^2}{\Lambda^2} \Big ] ,\\
    \beta_6 &\sim \frac{2}{v^4}
   \Big [ 
   z ( 5 M^2 + 3 M_A^2) - 2 M_A^2 \frac{g_1 v^2}{\Lambda^2} + 12 M_A^2 \frac{g_2 v^2}{\Lambda^2}
   \Big ] ,\\
   \beta_7 &\supset  \frac{4}{v^4}  ( 4 M^2 + 11 M_A^2  ) \Big (z - \frac{g_1 v^2}{\Lambda^2} \Big ) ,\\
   \beta_{8} &\supset - \frac{1}{6 v^2}
   \Big [  6 \frac{g_2 v^2}{\Lambda^2} (M^2 + 5 M_A^2) + M_A^2 \Big ( -z + \frac{g_1 v^2}{\Lambda^2} \Big ) \Big ] ,\\
   \beta_{10} &\sim -\frac{4}{v^4} \frac{g_2 v^2}{\Lambda^2}  (2 M^2 + M_A^2).
\end{align}
\label{beta.functs}
\end{subequations}
\endgroup

\section{One-loop mixing matrices}\label{sec.oneloop.mix}

We are now in a position to compare our results with those in the literature~\cite{Cheung:2015aba}. By inspecting Eq.(\ref{tilde.lambdas})  we obtain the mixing matrix represented in Table~\ref{tab.1}. We find agreement with the results of~\cite{Cheung:2015aba} with the exception of the mixing of $\phi^4 D^2$ operators with $F^2\phi^2$. More specifically, a closer inspection of Eq.~(\ref{tilde.lambdas}) shows that the operator 
\begin{align}
    {\cal I}_7 = \int \d \Big ( \phi^\dagger \phi - \frac{v^2}{2} \Big ) (D^\mu \phi)^\dagger D_\mu \phi,
\end{align} 
respects the mixing pattern derived in~\cite{Cheung:2015aba}, whereas the operator \begin{align}
    {\cal I}_6 = \int \d  \Big ( \phi^\dagger \phi - \frac{v^2}{2} \Big ) (\phi^\dagger D^2 \phi + \mathrm{h.c.} ),
\end{align} 
does not since it mixes with 
\begin{align}
    {\cal I}_{10} = \int \d 
F^2_{\mu\nu} \Big ( \phi^\dagger \phi - \frac{v^2}{2} \Big ).
\end{align}

\begin{table}[!t]
	$\begin{array}{*1c| *1c| *2c|}
	& \quad F^2 \phi^2\quad\mbox{} & \quad\phi^4D^2\quad\mbox{}  & \quad\phi^6\quad\mbox{}   \\
	\hline
	\rowcolor{zero2}
	\cw F^2 \phi^2 
	& \cw & & \\
	\hline
	\rowcolor{zero2}
	\cw \phi^4 D^2 
	& \cw \times & &  \\
	\rowcolor{white}
	\phi^6 
	&  &  & \\ \hline
	\end{array}$
	\caption{One-loop operator mixing matrix in the Abelian HK model. Shaded entries  denote a vanishing coefficient. The $\times$ indicates an entry that should vanish according to the non-renormalization theorem of~\cite{Cheung:2015aba} but that does not given the coefficients in Eq.~(\ref{tilde.lambdas}).}
	\label{tab.1}
\end{table}

There is an elegant cohomological interpretation of this result. One can find  ${\cal S}_0$-invariant combinations of gauge invariant operators that do not depend on the antifields, in very much the same way as in Eq.~(\ref{Y1.landau}). Notice that these invariants depend on $\sigma,\chi$ only via $\phi$ and they are generated by ${\cal Z}_2$ (now to be understood in the target theory). In particular one finds
\begin{align}
    {\cal S}_0 \int \d {\cal Z}_2 & = \int \d \Big ( \phi \frac{\delta S}{\delta \phi} + \phi^\dagger \frac{\delta S}{\delta \phi^\dagger} \Big ) \nonumber \\ & = 
    \int \d \Big [ - (D^2 \phi)^\dagger \phi - \phi^\dagger D^2 \phi  -2 \frac{M^2}{v^2} \Big ( \phi^\dagger \phi - \frac{v^2}{2} \Big ) \phi^\dagger \phi \Big ],
\end{align}
which is gauge-invariant. Thus {\it any} invariant of the form
\begin{align}
{\cal S}_0 \int \d Q(\phi,\phi^\dagger,A_\mu){\cal Z}_2, 
\label{Q.invs}
\end{align}
is gauge invariant if $Q$ is a gauge-invariant polynomial. Being cohomologically trivial, the above family of invariants can be added order by order in the loop expansion without changing the physical observables of the theory. Intuitively the simultaneous variation of  the coefficients of the operators entering in the invariants~(\ref{Q.invs}) cannot affect the physics since the variation is proportional to the equations of motion.

This is an example of the aforementioned fact that the mixing between gauge-invariant and alien operators (which are cohomologically trivial with respect to the linearized ST operator) can be made to vanish by a suitable basis choice in the space of local operators~\cite{Collins:1994ee,Joglekar:1975nu, KlubergStern:1974rs,KlubergStern:1974xv,KlubergStern:1975hc}.

This means that there is the freedom to replace the invariant ${\cal I}_6$ with the linear combination of ${\cal I}_2$ and ${\cal I}_3$ in Eq.~(\ref{inv.6}) up to a cohomologically trivial ${\cal S}_0$-invariant. This transformation induces the following shift on the space of the $\tilde{\lambda}$'s parameters:
\begin{align}
    \widetilde{\lambda}_2 &\rightarrow \widetilde{\lambda}_2 -  M^2 \widetilde{\lambda}_6;&
    \widetilde{\lambda}_3 &\rightarrow \widetilde{\lambda}_3 - \frac{2 M^2}{v^2} \widetilde{\lambda}_6.
\end{align}
For this new basis then, the non-renormalization theorem of~\cite{Cheung:2015aba} hold true.

In order to study the one-loop amplitudes dependence on the $g_i$'s and $z$ beyond the single higher-dimensional operator insertion approximation commonly used in the literature, we have reported in Table~\ref{tab.2} the dependence of the (shifted) $\tilde \lambda$'s coefficients on the $g_i$'s and $z$, based on the full one loop computation carried out in the present paper. 


The vanishing entries in Table~\ref{tab.2} can be partially understood in terms of the underlying amplitudes decomposition made transparent by the $X$-formalism. As explained above, the $\widetilde{\lambda}$'s are a linear combination of the $\lambda$'s coefficients multiplying gauge invariant operators which are independent from external sources of the $X$-theory, and of the coefficients $\vartheta, \theta$'s associated with invariants involving external sources insertions (the UV behaviour of which is more constrained than that of the fields). In particular, we find for the relevant operators in Table~\ref{tab.2} : 
\begin{align}
    \widetilde{\lambda}_4 &=  \lambda_4 + \frac{ g_1 \cfps{1}}{\Lambda^2};&
    \widetilde{\lambda}_5 &= \lambda_5;&
    \widetilde{\lambda}_8 &=  \lambda_8 + \frac{ g_2 \cfps{1}}{\Lambda^2};&
    \widetilde{\lambda}_9 &= \lambda_9.
    \label{1loop.full.cancs}
\end{align}

The $\cfps{1}$-terms can be neglected: they can only induce a $z$-dependence and thus do not contribute to the cancellations in Table \ref{tab.2}. Hence, the problem is reduced to the determination of the $g_i$'s dependence of the $\lambda$'s coefficients in the $X$-theory. One immediately sees that these coefficients cannot depend on $g_3$ since this is a trilinear vertex in $X_2$ that does not contribute to the 1-PI amplitudes of the starting theory at one loop. Thus, the last row of Table  \ref{tab.2} must hold, as  the only possible dependence on $g_3$ at one loop arises from the mapping to the target theory in Eq.(\ref{mapping}) and therefore governed by external amplitudes involving $\bar c^*$ and/or $R$ external sources, which do not enter in Eq.~(\ref{1loop.full.cancs}).

\begin{table}[!t]
\centering
    $\begin{array}{*1c| *9c *1c|}
    & ~~\widetilde{\lambda}_1~~ & ~~\widetilde{\lambda}_2~~ & ~~\widetilde{\lambda}_3~~ & ~~\widetilde{\lambda}_4~~ & ~~\widetilde{\lambda}_5~~ & ~~\widetilde{\lambda}_6~~ & ~~\widetilde{\lambda}_7~~ & ~~\widetilde{\lambda}_8~~ & ~~\widetilde{\lambda}_9~~ & ~~\widetilde{\lambda}_{10}~~  \\ [0.5ex] 
    \hline
    \rowcolor{white}
    z & & & & & & & & & & \\
    \rowcolor{zero2}
    \cw g_1 & \cw & \cw & \cw & \cw & \cw & \cw & \cw & \cw &  & \cw \\
    \rowcolor{zero2}
    \cw g_2 & \cw & \cw &  \cw & & & \cw & \cw & \cw & \cw & \cw \\
    \rowcolor{zero2}
    \cw g_3 & \cw & \cw & \cw & & & \cw & \cw & & & \cw \\
    \hline
\end{array}$
\caption{Dependence of the $\widetilde{\lambda}_i$'s on the higher dimensional coupling constants. Shaded entries denote that the dependence of the $\widetilde{\lambda}_i$ parameter on the corresponding coupling constant vanishes.}
\label{tab.2}
\end{table}
 
The remaining three forbidden dependences just seem to be an accidental consequence of the one-loop Feynman diagrams; as a result, cancellation patterns do not seem to lend themselves to an easy generalization to higher orders.

\section{Conclusions}\label{sec.concls}

In the present paper we have completed the investigation of the one-loop off-shell renormalization of  the Abelian Higgs-Kibble model supplemented at tree-level with all dimension 6 parity preserving on-shell inequivalent gauge-invariant operators. This was the last step towards the analysis of the SU(2)$\times$U(1) case.

We have shown that the $X$-theory formalism provides an effective way to work out the relevant GFRs, which in turn are found to have an ubiquitous effect on the one-loop UV coefficients of dimension 6 operators. In fact, since the GFRs are non linear and even non polynomial in the fields, it is advantageous to employ cohomological tools in order to disentangle the UV coefficients of the gauge-invariant operators from the spurious (and gauge-dependent) contributions associated with GFRs.

We have provided a full one-loop computation going beyond the customary linearized approximation in the higher dimensional couplings. All coefficients have been evaluated both in Feynman and in Landau gauge and the gauge independence of the UV coefficients of the gauge invariant operators explicitly checked. As expected, it does not hold unless  the effects of GFRs are properly accounted for.

We find that the pattern of operator mixing cancellations studied in the previous literature only holds off-shell if an appropriate choice of the on-shell equivalent operators is made. This can be traced back to the freedom of adding cohomologically trivial combinations of gauge-invariant operators at one loop order, thus selecting a particular basis of gauge-invariant on-shell inequivalent operators.

Application of the method presented to the SMEFT is currently under investigation.

\vfill\newpage
\appendix

\section{Functional Identities in the $X$-theory}\label{app:funct.ids}

\subsection{ST identities}

The ST identity (also known as the master equation in the BV approach) associated to the gauge group BRST symmetry reads
\begin{align}
	& {\cal S}(\G)  = \int \mathrm{d}^4x \, \Big [ 
	\partial_\mu \omega \frac{\delta \G}{\delta A_\mu} + \frac{\delta \G}{\delta \sigma^*} \frac{\delta \G}{\delta \sigma}  + \frac{\delta \G}{\delta \chi^*} \frac{\delta \G}{\delta \chi} 
	+ b \frac{\delta \G}{\delta \bar \omega} \Big ] = 0,
	\label{sti} 
\end{align}
or, at order $n$ in the loop expansion,
\begin{align}
    & {\cal S}(\G)^{(n)} = {\cal S}_0 (\G^{(n)}) +
    \sum_{j=1}^{n-1} 
    \Big (
    \frac{\delta \G^{(j)}}{\delta \sigma^*} \frac{\delta \G^{(n-j)}}{\delta \sigma}  
    + \frac{\delta \G^{(j)}}{\delta \chi^*} \frac{\delta \G^{(n-j)}}{\delta \chi}
    \Big ) = 0,
\end{align}
where ${\cal S}_0$ is the linearized ST operator:
\begin{align}
	{\cal S}_0  (\G^{(n)}) & = \int \!\mathrm{d}^4 x \, \Big [ \partial_\mu \omega \frac{\delta \G^{(n)}}{\delta A_\mu}  + 
	e\omega(\sigma+v)\frac{\delta \G^{(n)}}{\delta \chi}  
	-e\omega\chi\frac{\delta \G^{(n)}}{\delta \sigma}
+ b \frac{\delta \G^{(n)}}{\delta \bar \omega}  \nonumber \\
&  + \frac{\delta \G^{(0)}}{\delta \sigma} \frac{\delta \G^{(n)}}{\delta \sigma^*} + \frac{\delta \G^{(0)}}{\delta \chi} \frac{\delta \G^{(n)}}{\delta \chi^*} \Big ] \nonumber \\
& = s \G^{(n)} + \int \!\mathrm{d}^4 x \, \Big [ \frac{\delta \G^{(0)}}{\delta \sigma} \frac{\delta \G^{(n)}}{\delta \sigma^*} + \frac{\delta \G^{(0)}}{\delta \chi} \frac{\delta \G^{(n)}}{\delta \chi^*}  \Big ].
\label{S0}
\end{align}
${\cal S}_0$ maps the antifields $\sigma^*, \chi^*$ into the equations of motion of the fields $\sigma,\chi$, while it acts on the fields as the BRST operator $s$. Notice that, as explained before, we do not introduce an antifield for the gauge field $A_\mu$ since in the Abelian case treated here the gauge BRST transformation is linear.

The ST identity for the constraint BRST symmetry is
\begin{align}
    {\cal S}_{\scriptscriptstyle{C}}(\G) \equiv \int \!\mathrm{d}^4 x \, \Big [ v c \frac{\delta \G}{\delta X_1} 
 + \frac{\delta \G}{\delta \bar c^*}\frac{\delta \G}{\delta \bar c} \Big ] = 
 \int \!\mathrm{d}^4 x \, \Big [ v c \frac{\delta \G}{\delta X_1} 
 -(\square + m^2) c \frac{\delta \G}{\delta \bar c^*} \Big ] = 0,
 \label{sti.c} 
\end{align}
where in the latter equality we have used the fact that both the ghost $c$ and the antighost $\bar c$ are free:
\begin{align}
    \frac{\delta \G}{\delta \bar c} &= -(\square + m^2) c;&
    \frac{\delta \G}{\delta c} &= (\square + m^2) \bar c.   
    \label{c.eoms}
\end{align}

\subsection{$X_{1,2}$-equations}

By using Eq.~(\ref{c.eoms}) one sees that Eq.~(\ref{sti.c}) reduces to the $X_1$-equation of motion
\begin{align}
    \frac{\delta \G}{\delta X_1}=
    \frac{1}{v} (\square + m^2)
    \frac{\delta \G}{\delta \bar c^*}.
    \label{X1.eq}
\end{align}
Notice that this equation stays the same irrespectively of the presence of higher-dimensional gauge invariant operators added to the power-counting renormalizable action.

The $X_2$-equation is in turn given by
\begin{align}
	\frac{\delta \G}{\delta X_2} & =  \frac{1}{v} (\square + m^2) \frac{\delta \G}{\delta \bar c^*} + \frac{g_1 v}{ \Lambda^2} \frac{\delta \G}{\delta T_1} 
	+ \frac{g_2 v}{\Lambda^2} \frac{\delta \G}{\delta U} 
	+ \frac{g_3 v^3}{2 \Lambda^2} \frac{\delta \G}{\delta R} 
	- (\square + m^2)X_1\nonumber \\ 
	&- \Big [ (1+z) \square + M^2 \Big ] X_2 - v \bar c^* .
	\label{X2.eq}
\end{align}

\subsection{Solving the $X_{1,2}$-equations}

At order $n$, $n \geq 1$ in the loop expansion the $X_{1,2}$-equations reduce to
\begin{subequations}
\begin{align}
    \frac{\delta \G^{(n)}}{\delta X_1} &=
    \frac{1}{v} (\square + m^2)
    \frac{\delta \G^{(n)}}{\delta \bar c^*}, \\
    \frac{\delta \G^{(n)}}{\delta X_2} &=  \frac{1}{v} (\square + m^2) \frac{\delta \G^{(n)}}{\delta \bar c^*} + \frac{g_1 v}{ \Lambda^2} \frac{\delta \G^{(n)}}{\delta T_1} 
	+ \frac{g_2 v}{ \Lambda^2} \frac{\delta \G^{(n)}}{\delta U} 
	+ \frac{g_3 v^3}{2 \Lambda^2} \frac{\delta \G^{(n)}}{\delta R}.
\end{align}
	\label{nth.X12.eqs}
\end{subequations}
By using the chain rule for functional differentiation it is straightforward to see that Eqs.~(\ref{nth.X12.eqs}) entail that $\G^{(n)}$ only depends on the combinations:
\begin{subequations}
\begin{align}
	\tbarc &= \bar{c}^* + \frac{1}{v}(\square + m^2) (X_1 + X_2);& \tT &= T_1 + \frac{g_1 v}{\Lambda^2}X_2, \nonumber \\
	\tU &= U + \frac{g_2 v}{\Lambda^2}X_2;& \tR &= R + \frac{g_3 v^3}{2\Lambda^2}X_2.
    \label{X12.sols}
\end{align}
\end{subequations}

\section{The $b$- and the gauge ghost equation}\label{app.beq}

The set of the functional identities holding in the $X$-formulation of the Abelian HK model is completed by:
\begin{itemize}
\item The $b$-equation:
\begin{eqnarray}
	\frac{\delta \G}{\delta b} = \xi b - \partial A - \xi e v \chi ;
	\label{b.eq}
\end{eqnarray}
\item The antighost equation:
\begin{eqnarray}
	\frac{\delta \G}{\delta \bar \omega} = \square \omega + \xi e v \frac{\delta \G}{\delta \chi^*} .
	\label{antigh.eq}
\end{eqnarray} 
\end{itemize}

At orders $n\geq 1$
the $b$- and the antighost equations imply
\begin{align}
\frac{\delta \G^{(n)}}{\delta b} &= 0;&
\frac{\delta \G^{(n)}}{\delta \bar \omega} &=   \xi e v \frac{\delta \G^{(n)}}{\delta \chi^*},
\end{align}
so that at higher orders
the vertex functional does not depend on the
Nakanishi-Lautrup field $b$ and the dependence on the antighost is only via the combination
\begin{align}
    \widetilde{\chi}^* \equiv \chi^* + 
    \xi e v \bar \omega.
    \label{chistar.redef}
\end{align}

\section{Propagators}\label{app:propagators}

\subsection{The $X-\sigma$ sector}
Diagonalization of the quadratic part
of the action in this sector is achieved by setting
$$\sigma = \sigma' + X_1 + X_2.$$
Then one has
\begin{align}
    \Delta_{\sigma'\sigma'} &= \frac{i}{p^2 - m^2};&
    \Delta_{X_1 X_1} &= -\frac{i}{p^2 - m^2};&
    \Delta_{X_2X_2} = \frac{i}{(1+z) p^2 - M^2}.
    \label{scalar.props}
\end{align}

Several comments are in order here. At $g_1,g_2,g_3=0$ no higher dimensional interactions vertices are present. However, the model is still non power-counting renormalizabile, since the derivative interaction of the $X_{1,2}$-system $\sim (X_1 + X_2) \square (\phi^\dagger \phi)$ violates power-counting renormalizability as a consequence of the fact that the combination $X\equiv X_1+X_2$ has a propagator falling down as $1/p^2$ for large $p$ at $z \neq 0$, as can be seen from Eq.~(\ref{scalar.props}):
\begin{align}
  \Delta_{XX} = \Delta_{X_1X_1}+ \Delta_{X_2X_2} \sim -\frac{i z}{1+z} \frac{1}{p^2}.
\end{align}
On the other hand at $z=0$ $\Delta_{XX}$ goes as $1/p^4$ for large momenta and this compensates the two momenta from the $X \phi^\dagger \phi$ interaction vertex, giving rise to a power-counting renormalizable model (at zero $g_i$'s)~\cite{Quadri:2016wwl}.

\subsection{The gauge and ghost sector}

The diagonalization in the gauge sector is obtained by redefining the Nakanishi-Lautrup multiplier field 
\begin{align}
    b'= b - \frac{1}{\xi} \partial A - ev \chi .
\end{align}
Then, the $A_\mu$-propagator is
\begin{align}
    \Delta_{\mu\nu} = -i \Big ( \frac{1}{p^2-M_A^2} T_{\mu\nu} + \frac{1}{\frac{1}{\xi} p^2-M_A^2} \Big ); \qquad M_A = ev,
\end{align}
whereas the the Nakanishi-Lautrup, pseudo-Goldstone and ghost propagators are
\begin{align}
    \Delta_{b'b'} &= i \frac{1}{\xi};& \Delta_{\chi\chi} &= \frac{i}{p^2- \xi M_A};&
   \Delta_{\bar \omega \omega} &= \frac{i}{p^2- \xi M^2_A}.
\end{align}

As usual, $\xi=0$ corresponds to the Landau gauge, whereas $\xi=1$ is the Feynman gauge.

Finally, the ghost associated to the constraint BRST symmetry is free:
\begin{align}
    \Delta_{\bar c c} = \frac{-     i}{p^2-m^2}.
\end{align}

\newpage
\section{\label{app:g.inv.ops}List of Gauge-invariant Operators}
\subsection{Pure external sources invariants}
\begingroup
\allowdisplaybreaks
\begin{subequations}
\begin{align}
    & \vartheta_1 \int \d \bar c^*;& 
    & \vartheta_2 \int \d T_1; &
    & \vartheta_3 \int \d U; & 
    &  \vartheta_4 \int \d R, \\
    & \frac{\vartheta_5}{2} \int \d  (\bar c^*)^2; & &
      \frac{\vartheta_6}{2} \int \d  T_1^2; & 
    & \frac{\vartheta_7}{2} \int \d  U^2;& &
      \frac{\vartheta_8}{2} \int \d  R^2, \\
    & \frac{\vartheta_9}{2}  \int \d  \bar c^* \square \bar c^*;& &
    \frac{\vartheta_{10}}{2} \int \d T_1 \square T_1; &
    & \frac{\vartheta_{11}}{2} \int \d U \square U; & &
    \frac{\vartheta_{12}}{2} \int \d R \square R,\\
    & \vartheta_{13} \int \d \bar c^* T_1; & &
    \vartheta_{14} \int \d \bar c^* U;&
    & \vartheta_{15} \int \d \bar c^* R; & &
    \vartheta_{16} \int \d T_1 U,\\
    & \vartheta_{17} \int \d T_1 R ;& &
    \vartheta_{18} \int \d U R;&    
    & \vartheta_{19} \int \d \bar c^* \square T_1; & &
    \vartheta_{20} \int \d \bar c^* \square U,\\
    & \vartheta_{21} \int \d \bar c^* \square R; & &
    \vartheta_{22} \int \d T_1 \square U;&
    & \vartheta_{23} \int \d T_1 \square R;& &
    \vartheta_{24} \int \d U \square R,\\
    & \frac{\vartheta_{25}}{6} \int \d (\bar c^*)^3; & &
    \frac{\vartheta_{26}}{6} \int \d T_1^3;&
    & \frac{\vartheta_{27}}{6} \int \d U^3;& &
    \frac{\vartheta_{28}}{6} \int \d R^3,\\
    & \frac{\vartheta_{29}}{2} \int \d (\bar c^*)^2 T_1;& &
    \frac{\vartheta_{30}}{2} \int \d (\bar c^*)^2 U;&
    & \frac{\vartheta_{31}}{2} \int \d (\bar c^*)^2 R;& &
    \frac{\vartheta_{32}}{2} \int \d \bar c^* T_1^2,\\
    & \frac{\vartheta_{33}}{2} \int \d \bar c^* U^2;& &
    \frac{\vartheta_{34}}{2} \int \d \bar c^* R^2;&
    & \vartheta_{35} \int \d \bar c^* T_1 U ;& &
    \vartheta_{36} \int \d \bar c^* T_1 R,\\
    & \vartheta_{37} \int \d \bar c^* U R;& &
    \frac{\vartheta_{38}}{2} \int \d T_1^2 U;&
    & \frac{\vartheta_{39}}{2} \int \d T_1^2 R;& &
    \frac{\vartheta_{40}}{2} \int \d T_1 U^2,\\
    & \frac{\vartheta_{41}}{2} \int \d T_1 R^2;& &
    \vartheta_{42} \int \d T_1 U R;&    
    & \frac{\vartheta_{43}}{2} \int \d U^2 R; & &
    \frac{\vartheta_{44}}{2} \int \d U R^2.
\end{align}
 \label{pure-ext.srcs}
\end{subequations}
\endgroup

\subsection{Mixed field-external sources invariants}
\begingroup
\allowdisplaybreaks
\begin{subequations}
\begin{align}
    & \theta_1 \int \d \bar c^* \Big ( \phi^\dagger \phi - \frac{v^2}{2} \Big ); &
    &  \theta_2 \int \d T_1 \Big ( \phi^\dagger \phi - \frac{v^2}{2} \Big ); &
    & \theta_3 \int \d U \Big ( \phi^\dagger \phi - \frac{v^2}{2} \Big ), \\
    & \theta_4 \int \d R \Big ( \phi^\dagger \phi - \frac{v^2}{2} \Big );& 
    & \theta_5 \int \d \bar c^* (D^\mu \phi)^\dagger D_\mu \phi;& 
    &  \theta_6 \int \d T_1 (D^\mu \phi)^\dagger D_\mu \phi, \\
    & \theta_7 \int \d U (D^\mu \phi)^\dagger D_\mu \phi; & & 
      \theta_8 \int \d R (D^\mu \phi)^\dagger D_\mu \phi; &
    & \theta_9 \int \d \bar c^* \Big ( \phi^\dagger D^2\phi + \mathrm{h.c.} \Big ), \\
    &  \theta_{10} \int \d T_1 \Big ( \phi^\dagger D^2\phi + h.c. \Big ); & 
    & \theta_{11} \int \d U \Big ( \phi^\dagger D^2\phi + \mathrm{h.c.} \Big ); & & 
      \theta_{12} \int \d R \Big ( \phi^\dagger D^2\phi + \mathrm{h.c.} \Big ), \\
    & \frac{\theta_{13}}{2} \int \d \bar c^* \Big ( \phi^\dagger \phi - \frac{v^2}{2} \Big )^2;& & 
      \frac{\theta_{14}}{2} \int \d T_1 \Big ( \phi^\dagger \phi - \frac{v^2}{2} \Big )^2; &
    & \frac{\theta_{15}}{2} \int \d U \Big ( \phi^\dagger \phi - \frac{v^2}{2} \Big )^2,\\
    & \frac{\theta_{16}}{2} \int \d R \Big ( \phi^\dagger \phi - \frac{v^2}{2} \Big )^2; &
    & \theta_{17} \int \d \bar c^* F_{\mu\nu}^2; & & 
      \theta_{18} \int \d T_1 F_{\mu\nu}^2, \\
    & \theta_{19} \int \d U F_{\mu\nu}^2;& & 
      \theta_{20} \int \d R F_{\mu\nu}^2;&
    & \frac{\theta_{21}}{2} \int \d (\bar c^*)^2 \Big ( \phi^\dagger \phi - \frac{v^2}{2} \Big ), \\
    & \frac{\theta_{22}}{2} \int \d T_1^2 \Big ( \phi^\dagger \phi - \frac{v^2}{2} \Big );&
    & \frac{\theta_{23}}{2} \int \d U^2 \Big ( \phi^\dagger \phi - \frac{v^2}{2} \Big ); & & 
      \frac{\theta_{24}}{2} \int \d R^2 \Big ( \phi^\dagger \phi - \frac{v^2}{2} \Big ), \\
   & \theta_{25} \int \d \bar c^* T_1 \Big ( \phi^\dagger \phi - \frac{v^2}{2} \Big ); & & 
    \theta_{26} \int \d \bar c^* U \Big ( \phi^\dagger \phi - \frac{v^2}{2} \Big ); &
   & \theta_{27} \int \d \bar c^* R \Big ( \phi^\dagger \phi - \frac{v^2}{2} \Big ), \\
   & \theta_{28} \int \d  T_1 U \Big ( \phi^\dagger \phi - \frac{v^2}{2} \Big );&
   & \theta_{29} \int \d  T_1 R \Big ( \phi^\dagger \phi - \frac{v^2}{2} \Big ); & & 
 	 \theta_{30}  \int \d  U R \Big ( \phi^\dagger \phi - \frac{v^2}{2} \Big ). 
 \label{mixed.field-ext.srcs}
\end{align}
\end{subequations}
\endgroup

\subsection{Gauge invariants depending only on the fields}
\begin{subequations}
\begin{align}
    & \lambda_1 \int \d \Big ( \phi^\dagger \phi - \frac{v^2}{2} \Big ) ;
    & & 
    \lambda_2 \int \d \Big ( \phi^\dagger \phi - \frac{v^2}{2} \Big )^2 , \\
    & \lambda_3 \int \d \Big ( \phi^\dagger \phi - \frac{v^2}{2} \Big )^3;
    & & 
    \lambda_4 \int \d (D^\mu \phi)^\dagger D_\mu \phi,\\
    & \lambda_5 \int \d \Big ( \phi^\dagger D^{(\mu\nu\mu\nu)} \phi +\mathrm{h.c.} \Big ); 
    & & \lambda_6 \int \d \Big ( \phi^\dagger \phi - \frac{v^2}{2} \Big )
    ( \phi^\dagger D^2 \phi + \mathrm{h.c.} ),\\
   & \lambda_7 \int \d \Big ( \phi^\dagger \phi - \frac{v^2}{2} \Big )
    (D^\mu \phi)^\dagger D_\mu \phi; &  &
    \lambda_8 \int \d F_{\mu \nu}^2 ,\\
   & \lambda_9 \int \d \partial^\rho F_{\rho \mu} 
   \partial_\sigma F^{\sigma \mu}; & & 
   \lambda_{10} \int \d F_{\mu \nu}^2 \Big ( \phi^\dagger \phi - \frac{v^2}{2} \Big ),   
   \end{align}
\label{ops.only.fields}
\end{subequations}
where $D^{(\mu\nu\mu\nu)}$ denotes complete symmetrization over $\mu,\nu$:
\begin{align}
    D^{(\mu\nu\mu\nu)} \phi \equiv 
    [(D^2)^2 + D^\mu D^\nu D_\mu D_\nu + D^\mu D^2 D_\mu ] \phi .
\end{align}
Notice that in the text we have denoted by $\lop{j}$ the invariant with coefficient $\lambda_j$.

\section{On-shell Reduction of dim.6 Field-Dependent Gauge Invariant Operators}\label{app:reduction}

We consider in this Appendix the on-shell reduction of dimension 6 operators
in the target theory. The relevant classical gauge-invariant action $S$ is obtained from the first four lines of Eq.(\ref{tree.level})  by going on-shell with $X_{1,2}$.

The corresponding equations of motion for the gauge field and the scalar $\phi$ are
\begin{subequations}
\begin{align}
    \frac{\delta S}{\delta A_\mu} &=
    \partial^\rho F_{\rho \mu} + i 
    \Big [ \phi^\dagger D_\mu \phi -
    (D_\mu \phi)^\dagger \phi \Big ], 
     \\
    \frac{\delta S}{\delta \phi} &= 
    - (D^2 \phi)^\dagger - \frac{M^2}{v^2}
    \Big ( \phi^\dagger \phi - 
    \frac{v^2}{2} \Big ) \phi^\dagger ,
     \\
    \frac{\delta S}{\delta \phi^\dagger} &= 
    - (D^2 \phi) - \frac{M^2}{v^2}
    \Big ( \phi^\dagger \phi - 
    \frac{v^2}{2} \Big ) \phi.
    \end{align}	
\label{tree.eoms}
\end{subequations}
Since we will be interested only in the one-loop corrections that are linear in the $g_i$'s and $z$ we can limit ourselves to the leading order equations of motion in Eq.~(\ref{tree.eoms}); also we recall here the identity 
\begin{align}
	[D_\mu , D_\nu] = -i F_{\mu\nu}.
	\label{comm}	
\end{align}

The on-shell independent dimension 6 operators can be chosen to be $\lop{3},\lop{7}$ and $\lop{10}$.
Notice that the operator in the tree-level vertex functional
\begin{align}
    \int \d 
     \Big ( \phi^\dagger \phi - \frac{v^2}{2} \Big )\square
     \Big ( \phi^\dagger \phi - \frac{v^2}{2} \Big ) & = \int \d 
     \Big ( \phi^\dagger \phi - \frac{v^2}{2} \Big ) \Big [
     (D^2\phi)^\dagger \phi + \phi^\dagger (D^2\phi) + 
     2 (D^\mu \phi)^\dagger D_\mu \phi \Big ], 
\end{align}
can be represented in terms of invariants in the contractible pairs basis as in the r.h.s. of the above equation. Therefore we just need to reduce all $\lop{i}$'s invariants in terms of $\lop{3},\lop{7}$ and $\lop{10}$ by using the equations of motion (\ref{tree.eoms}).

Let us start from $\lop{5}$. This operator contains three terms,
namely: 
\begin{align}
	&\int \d \phi^\dagger D^4 \phi;& & \int \d \phi^\dagger D^\mu D^2 D_\mu \phi;& \int \d \phi^\dagger D^\mu D^\nu D_\mu D_\nu \phi.
	\label{3terms}
\end{align}
Then one finds that:
\begin{itemize}
    \item 
    Integration by parts gives:
    \begin{align}
        \int \d \phi^\dagger (D^2)^2 \phi = \int \d (D^2\phi)^\dagger D^2\phi \sim
        \int \d \frac{M^4}{v^4} \Big ( \phi^\dagger \phi - \frac{v^2}{2} \Big )^2 \phi^\dagger \phi
    \end{align}
    where the equations of motion for $\phi,\phi^\dagger$ have been used in the last line. Hence we obtain
    \begin{align}
        \int \d \phi^\dagger (D^2)^2 \phi \sim 
        \int \d \Big \{ \frac{M^4}{v^4} \Big ( \phi^\dagger \phi - \frac{v^2}{2} \Big )^3  + \frac{M^4}{2 v^2} \Big ( \phi^\dagger \phi - \frac{v^2}{2} \Big )^2 \Big \} =
        \frac{M^4}{2v^2}\lop{2} +
        \frac{M^4}{v^4} \lop{3} .
        \label{inv5.1}
    \end{align}
    \item The second term can be rewritten as follows
    \begin{align}
        \int \d \phi^\dagger D^\mu D^2 D_\mu \phi &=
        \int \d \phi^\dagger D^\mu D^\rho D_\mu D_\rho \phi + 
        \int \d \phi^\dagger D^\mu D^\rho [ D_\rho, D_\mu] \phi \nonumber \\
        & = \int \d \phi^\dagger D^\mu D^\rho D_\mu D_\rho \phi  
        - i \int \d (D^\rho D^\mu \phi)^\dagger  F_{\rho\mu} \phi, 
    \end{align}
    where in the last line we have used Eq.~(\ref{comm}) and integrated by parts.
Now
\begin{align*}
    - i \int \d (D^\rho D^\mu \phi)^\dagger  F_{\rho\mu} \phi  & =
    - \frac{i}{2} \int \d ([D^\rho, D^\mu] \phi)^\dagger  F_{\rho\mu} \phi 
 =
    - \frac{1}{2} \int \d  F^2_{\rho\mu} \phi^\dagger   \phi,
    \label{inv5.2}
\end{align*}
again by using Eq.~(\ref{comm}). Eventually we arrive at the result
    \begin{align}
        \int \d \phi^\dagger D^\mu D^2 D_\mu \phi &=
        \int \d \phi^\dagger D^\mu D^\rho D_\mu D_\rho \phi  ~
        - \frac{1}{2} \int \d  F^2_{\rho\mu} \Big ( \phi^\dagger \phi - \frac{v^2}{2} \Big )  -  \frac{v^2}{4} \int \d F^2_{\rho\mu} \nonumber \\
        & =  \int \d \phi^\dagger D^\mu D^\rho D_\mu D_\rho \phi
        ~  -  \frac{v^2}{4}
        \lop{8} -  
        \frac{1}{2} \lop{10}  .
    \end{align}
\item 
We are finally left with the decomposition of the last term in 
\1eq{3terms}. One has
\begin{align}
    \int \d \phi^\dagger D^\mu D^\rho D_\mu D_\rho \phi & = 
    \int \d \Big [ \phi^\dagger D^4 \phi + \phi^\dagger D^\mu [ D^\rho, D_\mu] D_\rho \phi \Big ] \nonumber \\
    & = \int \d \Big [ (D^2 \phi)^\dagger D^2 \phi - i F_{\mu \rho} (D^\mu \phi)^\dagger D^\rho \phi \Big ],
\end{align}
where we have used Eq.~(\ref{comm}) and integrated by parts.
It is convenient to split the last term in the above equation as follows
\begin{align}
    i \int \d F_{\mu \rho} (D^\mu \phi)^\dagger D^\rho \phi & =
    \frac{i}{2} \int \d F_{\mu \rho} \Big \{ (D^\mu \phi)^\dagger D^\rho \phi +
    (D^\mu \phi)^\dagger D^\rho \phi \} \nonumber \\
    & = \int \d \Big \{ -\frac{i}{2} \partial^\rho F_{\rho \mu} \Big [ 
    \phi^\dagger D^\mu \phi - (D^\mu \phi)^\dagger \phi \Big ] \nonumber \\
    & \qquad \qquad - \frac{i}{4}
   F_{\mu\rho} \Big [ \phi^\dagger [D^{\mu}, D^\rho] \phi +  ([ D^\rho, D^\mu] \phi)^\dagger \phi \Big ] \Big \} \nonumber \\
    & = \int \d \Big \{ -\frac{i}{2} \partial^\rho F_{\rho \mu} \Big [ 
    \phi^\dagger D^\mu \phi - (D^\mu \phi)^\dagger \phi \Big ]
    -\frac{1}{2} F^2_{\mu\rho} \phi^\dagger \phi \Big \}.
\end{align}
By using the $A_\mu$-equation of motion (\ref{tree.eoms}) the first term in the last line of the above equation becomes
\begin{align}
    -\frac{i}{2} \int \d & \partial^\rho F_{\rho \mu} \Big [ 
    \phi^\dagger D^\mu \phi - (D^\mu \phi)^\dagger \phi \Big ]   \sim
    \nonumber \\
    & - \frac{1}{2}
    \int \d 
    \Big ( \phi^\dagger D^\mu \phi - 
    (D^\mu \phi)^\dagger \phi  \Big ) 
    \Big ( \phi^\dagger D_\mu \phi - 
    (D_\mu \phi)^\dagger \phi \Big ) \nonumber \\
    & = \int \d \Big \{ 
    \phi^\dagger \phi ~ (D^\mu \phi)^\dagger D_\mu \phi - \frac{1}{2}
    \Big [ 
    \phi^\dagger D_\mu \phi ~\phi^\dagger D^\mu \phi + \mathrm{h.c.}
    \Big ]
    \Big \}.
\end{align}
Integrating by parts the last term in the last line of the above equation one finds
\begin{align}
   - \frac{1}{2} \int \d   \Big [ 
    \phi^\dagger D_\mu \phi ~\phi^\dagger D^\mu \phi + \mathrm{h.c.}
    \Big ] =
    \int \d \Big \{ 2 \phi^\dagger \phi (D^\mu \phi)^\dagger D_\mu \phi + \frac{1}{2}
    \phi^\dagger \phi \Big [ \phi^\dagger D^2 \phi + (D^2 \phi)^\dagger \phi \Big ] \Big \},
\end{align}
and thus
\begin{align}
    &-\frac{i}{2} \int \d \partial^\rho F_{\rho \mu} \Big [ 
    \phi^\dagger D^\mu \phi - (D^\mu \phi)^\dagger \phi \Big ]   \sim
    \nonumber \\
     & - \frac{1}{2}
    \int \d 
    \Big ( \phi^\dagger D^\mu \phi - 
    (D^\mu \phi)^\dagger \phi  \Big ) 
    \Big ( \phi^\dagger D_\mu \phi - 
    (D_\mu \phi)^\dagger \phi \Big ) \nonumber \\
    & = \int \d \Big \{ 
    3 \phi^\dagger \phi ~ (D^\mu \phi)^\dagger D_\mu \phi + \frac{1}{2} 
     \phi^\dagger \phi \Big [ \phi^\dagger D^2 \phi + (D^2 \phi)^\dagger \phi \Big ]
     \Big \}.
\end{align}
Putting everything together we find
\begin{align}
      &\int \d \phi^\dagger D^\mu D^\rho D_\mu D_\rho \phi  \sim 
    \nonumber \\
    & = \int \d \Big \{ (D^2 \phi)^\dagger D^2 \phi 
    -3 \phi^\dagger \phi ~ (D^\mu \phi)^\dagger D_\mu \phi -
    \frac{1}{2} \phi^\dagger \phi \Big [ \phi^\dagger D^2 \phi + (D^2 \phi)^\dagger \phi \Big ]
    + \frac{1}{2} F_{\mu\rho}^2 \phi^\dagger \phi \Big \} \nonumber \\
    & \sim \int \d
    \Big \{ \frac{M^2}{v^2}
    \Big ( 1 + \frac{M^2}{v^2}
    \Big ) \Big ( \phi^\dagger \phi - \frac{v^2}{2} \Big )^3 - 3 \Big ( \phi^\dagger \phi - \frac{v^2}{2} \Big ) (D^\mu \phi)^\dagger D_\mu \phi 
    +\frac{1}{2} F_{\mu\rho}^2 
    \Big ( \phi^\dagger \phi - \frac{v^2}{2} \Big )
    \nonumber \\
    & 
    \qquad \qquad 
    - \frac{3v^2 }{2} (D^\mu \phi)^\dagger D_\mu \phi 
    +  M^2  \Big ( 1 + \frac{M^ 2}{2 v^2} \Big )   \Big ( \phi^\dagger \phi - \frac{v^2}{2} \Big )^2 +
    \frac{1}{4} M^2 v^2 \Big ( \phi^\dagger \phi - \frac{v^2}{2} \Big )
    + \frac{v^2}{4} F_{\mu\rho}^2
   \Big \} \nonumber \\
   &
   = 
    \frac{1}{4} M^2 v^2 \lop{1}
    + M^2  \Big ( 1 + \frac{M^ 2}{2 v^2} \Big ) 
    \lop{2} +
   \frac{M^2}{v^2}
    \Big ( 1 + \frac{M^2}{v^2}
    \Big ) \lop{3} 
    - \frac{3}{2}v^2 \lop{4}
    - 3 \lop{7}
    + \frac{v^2}{4} \lop{8}
    + \frac{1}{2} \lop{10}.
    \label{inv5.3}
\end{align}
\end{itemize}
By using Eqs.(\ref{inv5.1}), (\ref{inv5.2}) and (\ref{inv5.3}) we obtain
\begin{align}
    \lop{5} \sim
     \frac{1}{2} M^2 v^2 \lop{1} +
    M^2
    \Big ( 
    2  + \frac{3}{2}\frac{M^2}{v^2} 
    \Big ) 
    \lop{2} 
    + \frac{M^2}{v^2}
    \Big ( 2 + \frac{3M^2}{v^2}
    \Big ) \lop{3}
    - 3 v^2 \lop{4}
    - 6 \lop{7}  
    + \frac{1}{4} v^2 \lop{8}
    +\frac{1}{2} \lop{10}.  
\end{align}

We now move to $\lop{6}$. By using the equations of motion for $\phi,\phi^\dagger$ in Eq.(\ref{tree.eoms}) we find
\begin{align}
    \lop{6} \sim \int \d \Big \{
    - 2 \frac{M^2}{v^2} 
    \Big ( \phi^\dagger \phi - \frac{v^2}{2} \Big )^3 -
    M^2 \Big ( \phi^\dagger \phi - \frac{v^2}{2} \Big )^2
    \Big \} = - 2 \frac{M^2}{v^2} \lop{3} - M^2 \lop{2}.
    \label{inv.6}
\end{align}

Finally we need to consider $\lop{9}$.
Use of $A_\mu$-equation of motion yields
\begin{align}
    \lop{9} & \sim -
    \int \d \Big [ \phi^\dagger D_\mu \phi - (D_\mu \phi)^\dagger \phi \Big  ]^2 \sim
    \int \d \phi^\dagger \phi \Big [ 6 (D^\mu \phi)^\dagger D_\mu \phi + \phi^\dagger D^2 \phi + (D^2\phi)^\dagger \phi \Big ] \nonumber \\
    & = \int \d \Big [ \Big ( \phi^\dagger \phi - \frac{v^2}{2} \Big ) + \frac{v^2}{2} \Big ] \Big \{ 6 (D^\mu \phi)^\dagger D_\mu \phi  - 2 \frac{M^2}{v^2} \Big ( \phi^\dagger \phi - \frac{v^2}{2} \Big ) ^2 -
    M^2 \Big ( \phi^\dagger \phi - \frac{v^2}{2} \Big ) \Big \}
    \nonumber \\
    & =
    - \frac{M^2 v^2}{2} \lop{1}
    - 2 M^2 \lop{2}
    - 2 \frac{M^2}{v^2} \lop{3}
    + 3 v^2 \lop{4} +
    6 \lop{7}.
\end{align}


%

\end{document}